\documentclass[review,12pt]{article}
\usepackage[utf8]{inputenc}
\usepackage[english]{babel}
\usepackage[T1]{fontenc}

\usepackage{authblk}
\usepackage{csquotes}
\usepackage{dirtytalk}
\usepackage{lipsum}
\usepackage{graphicx}
\usepackage{tikz}
\usepackage{setspace}
\usepackage{placeins}

\usepackage{amsmath}
\usepackage{amsfonts}
\usepackage{amssymb}
\usepackage{amsthm}
\usepackage{verbatim}
\usepackage{bm}
\usepackage{mathtools}
\usepackage{tcolorbox}
\usepackage{outlines}
\usepackage{makecell}
\usepackage{multirow}
\usepackage{hyperref}
\usepackage{colortbl}
\usepackage{natbib}
\usepackage{orcidlink}

\hypersetup{
    colorlinks=true,
    linkcolor=blue,
    filecolor=blue,
    urlcolor=blue,
    citecolor=blue
}
\newcommand{\eg}{\emph{e.g.}}
\newcommand{\Eg}{\emph{E.g.}}

\newcommand{\brp}[1]{\left( #1 \right)}

\newcommand{\brc}[1]{\left\{ #1 \right\}}

\newcommand{\abs}[1]{\lvert #1 \rvert}

\newcommand{\norm}{\mathcal{N}} 

\newcommand{\bigO}{\operatorname{\mathcal{O}}}

\newcommand{\I}{\mathbb{I}} 
\newcommand{\F}{\mathcal{F}} 
\newcommand{\td}{\mathit{t}}

\DeclarePairedDelimiter\ceil{\lceil}{\rceil}
\DeclarePairedDelimiter\floor{\lfloor}{\rfloor}

\usepackage[margin=1in]{geometry}

\newcommand{\SRD}[0]{\operatorname{SRD}}
\newcommand{\SRDnorm}[0]{\operatorname{\underline{SRD}}}

\begin{document}

\title{Testing Rankings with Cross-Validation}
\date{}

\author[1,2]{Balázs R. Sziklai
\texorpdfstring{%
\orcidlink{0000-0002-0068-8920}
}{}}
\affil[1]{%
Centre for Economic and Regional Studies, Budapest, Hungary}
\affil[2]{%
Department of Operations Research and Actuarial Sciences, Corvinus University of Budapest}

\author[3]{Máté Baranyi
\texorpdfstring{%
\orcidlink{0000-0003-4415-4805}
}{}}
\affil[3]{%
Department of Stochastics, Budapest University of Technology and Economics, Budapest, Hungary}

\author[4]{Károly Héberger
\texorpdfstring{%
\orcidlink{0000-0003-0965-939X}
}{}}
\affil[4]{
Plasme Chemistry Reasearch Group, Institute of Materials and Environmental Chemistry,\\ ELKH Research Centre for Natural Sciences, Budapest, Hungary}

\maketitle

\begin{abstract}
This research investigates how to determine whether two rankings come from the same distribution. We evaluate three hybrid tests: Wilcoxon’s, Dietterich’s, and Alpaydin’s statistical tests combined with cross-validation (CV), each operating with folds ranging from 5 to 10, thus altogether 18 variants. We have applied these tests in the framework of a popular comparative statistical test, the Sum of Ranking Differences that builds upon the Manhattan distance between the rankings.
The introduced methodology is widely applicable from machine learning through social sciences.
To compare these methods, we have followed an innovative approach borrowed from Economics. We designed nine scenarios for testing type I and II errors. These represent typical situations (that is, different data structures) that CV tests face routinely. The optimal CV method depends on the preferences regarding the minimization of type I/II errors, size of the input, and expected patterns in the data. The Wilcoxon method with eight folds proved to be the best for all three investigated input sizes. Although the Dietterich and Alpaydin methods are the best in type I situations, they fail badly in type II cases. We demonstrate our results on real-world data, borrowed from chess and chemistry. Overall we cannot recommend either Alpaydin or Dietterich as an alternative to Wilcoxon cross-validation.
\end{abstract}


\section{Introduction}

Ranking objects is one of the most commonly applied computational tasks. In social sciences, universities are ranked by excellence, sports teams are ranked according to various performance measures, politicians are ranked by their popularity, \emph{etc.} In machine learning, query results are ranked by search engines, features are ranked during feature selection, algorithms are ranked by their performance \emph{etc.} Yet, there are still unresolved tasks regarding the statistics of rankings. One such task is to attribute uncertainty to ranks. This is an extremely difficult problem as the underlying numerical scale contains many variance elements (bias, random errors, precision, accuracy, ruggedness, repeatability, reproducibility, \emph{etc.}). It would be expedient to calculate the standard deviation of ranks, although it can be done in at least two ways: in terms of objectives/solutions and of periods/use cases. For instance, a university ranking is based on various objectives (number of published papers, student/staff ratio, \emph{etc.}) and is different with regard to each objective. Furthermore, the ranking in terms of a single objective is uncertain as it changes each year.
In addition, aggregation itself is riddled with various paradoxes (Condorcet paradox, Arrow's and  Gibbard--Satterthwaite's impossibility theorem, discursive dilemma, \emph{etc.}) \citep[see][]{moulinHandbookComputationalSocial2016}.
Hence, it is no wonder that the scientists do not bother with the error of ranks but calculate them on the numerical scale.
The uncertainties in a set of alternative rankings, or in the raw data itself (from where the rankings are generated) are moderately studied, see Section~\ref{sec:litera}.

Another closely related problem is the comparison of rankings. How can we determine that two rankings are generated from the same distribution?
One way to answer this question is by using cross-validation (CV) techniques: looking at the subsets of the positions of two rankings, measuring their similarity, and comparing the obtained values for different subsets.

A unique contribution of this paper is in comparing the efficacy of CV methods in a ranking environment. For the latter, we have used the framework of Sum of Ranking Differences (SRD) – a popular comparative statistical test. The practical implication of this choice is that the difference between the rankings is measured in Manhattan distance or L1 norm.


SRD, introduced by \citet{hebergerSumRankingDifferences2010}, is a relatively novel test that is especially suitable for comparing methods in multi-objective optimization environments. In recent years, there have been many published papers, with topics ranging from machine learning~\citep{Moorthy2017},
through multi-criteria decision-making~\citep{GERE2021} and  pharmacology~\citep{Vajna2012},
to
political science~\citep{sziklaiApportionmentDistrictingSum2020}, and
even sports~\citep{West2018}, that apply or further extend the technique, showing its versatility.

SRD scores are computed by converting the input data into rankings, and then calculating the Manhattan distance  (Spearman's footrule if no ties are present) of each method's ranking and the reference ranking. The latter can be an external gold standard or an aggregate from the data \citep[see][]{hebergerSumRankingDifferences2010}.
The SRD algorithm contains two validation steps:
\begin{outline}
\1 In the so-called permutation test, we compare the methods' scores to the SRD scores of random rankings. For a detailed overview of this method see the article of~\cite{hebergerSumRankingDifferences2011}.
\1 In the second step, we cross-validate the results by re-sampling the data. We then assign uncertainties to the SRD scores, which are deterministic by nature, using cross-validation (CV). It also allows us to group and compare the methods differently than the pure SRD scores.
\end{outline}

SRD scores can be seen as a metric of the mean absolute error (on a rank scale) to the reference, which is the gold standard. It is easy to draw a parallel with machine learning, where the CV of the chosen error metric is a standard technique.
There are a lot of approaches in the literature about how to make conclusions from the many error scores calculated on different subsamples of the data. Not all of these can be easily adapted to ranking frameworks.

Originally, by \citet{KOLLARHUNEK2013139}, the Wilcoxon signed-rank test was proposed for CV purposes. Dietterich's CV $\td$-test~\citep{dietterichApproximateStatisticalTests1998} and Alpaydin's CV $\F$-test~\citep{alpaydinCombinedCvTest1999} are popular regarding cross-validated machine learning algorithms. An original contribution of this paper is the adaptation of these methods to the ranking framework and testing their performance. We have also proposed several scenarios for assessing type I and II errors, see Section~\ref{sec:scenarios}.
Based on these scenarios, practitioners can decide which CV algorithm is more suitable for their use case.
The scenario analysis reveals a mixed picture. Different CV methods prevail under different circumstances (type I/II errors, input size, data structure, \emph{etc.}). The Wilcoxon method with seven folds seems to be the best compromise, but its 8-fold version is also a viable alternative.

\section{Literature Overview}\label{sec:litera}

Solutions that employ ranking techniques are very common in statistical data analysis, pattern recognition, and machine learning in general. 
Applications are ranging from the ranking of image sets \citep{pedronetteImageRerankingRank2013} through feature selection \citep{ALAIZRODRIGUEZ2020105745} 
and matching \citep{jiangRankingListPreservation2021a} to image-text matching \citep{NIU2020107351} and image retrieval \citep{PEDRONETTE2017478}.

Rankings are evaluated and compared by various measures, among which Spearman's footrule (Manhattan distance among rankings) is quite common \citep{jiangRankingListPreservation2021a}. 
Note that SRD is a generalization of Spearman's footrule. For rankings -- that stem from partial orderings -- the $k$-nearest neighbor distance is often used, which in turn is a modified version of Kendall's tau measure \citep{BRANDENBURG2013}. 
\citet{LIU2021LOR} argues that in group decision making ordinal information is more flexible and intuitive for experts.

Sometimes, rankings are modeled in a parametric way built on top of one of the many available distance measures, \eg, Spearman's footrule; and these models are further generalized and combined with different ones.
Probably the most well-known is Mallows' ranking model \citep{mallowsNONNULLRANKINGMODELS1957}. It is studied intensively, for example, recently \citet{vitelliProbabilisticPreferenceLearning2017} introduced a new tractable method for Bayesian inference in Mallows' models.
There are many different approaches as well.
\citet{svendovaNovelMethodEstimating2017} proposes an indirect inference approach to estimate the latent signal parameters that might be causal for a set of observed rankings obtained from several assessors.
\citet{leeMixturesWeightedDistancebased2012} formulate models by considering weighted distances which allow different weights for different ranks.
In their earlier paper \citep{leeDistancebasedTreeModels2010} they combine a tree model and the existing distance-based models to build a model that can handle more complexity.
\citet{xuAnglebasedModelsRanking2018} propose a general angle-based model for ranking data, of which a distance-based model with Spearman's distance can be seen as a special case.
\citet{negahbanLearningComparisonsChoices2018} focuses on the multinomial logit model for the representation learning of rankings in a purely probabilistic sense, instead of relying on distance metrics.

Feature selection is a problem of high-dimensional data analysis, where the selection is usually based on some ranking of the features.
The feature vectors themselves are also frequently transformed into rankings.
\citet{chanFullRankingLocal2015} assigns rankings to picture pixels based on, \eg, the intensities of neighborhood pixels; and then find the closest \say{visual word} to each pixel. These \say{visual words} are prototype rankings, which resonate well with the reference ranking of the SRD framework as well. 
In Content-based Image Retrieval (CBIR) systems, ranking a collection of images is crucial, \eg, a ranked list of images is created based on a query image.
\citet{pedronetteImageRerankingRank2013} deal with re-ranking of the images based on additional contextual information, and define similarity of images based on the distance of their ranked lists (of the other images). Different re-rankings may be compared with one of the techniques introduced in our paper as well, \eg, by taking the original ranking as a reference.
\citet{jiangRankingListPreservation2021a} incorporate rankings in feature matching of images. They define the similarity between feature vectors of two images as the average top-K similarity (weighted Spearman's footrule) of 2-2 different rankings of their features; one ranking is calculated based on Euclidean distance, and the other one using angle correlation. They also say that the Euclidean distance based ranking list generation is sensitive to noise, which underlies the necessity of the study of ranking uncertainties.
\citet{zhaoNovelActiveLearning2019} view the sample selection problem of Active Learning (AL) methods as a ranking problem, and use weighted rank aggregation (they review a plethora of these) in their multiple query criteria AL-method. A comparison of rankings based on single queries would be possible by methods of our paper, \eg, by using the aggregated ranking as a reference.

\paragraph{Cross-validation}
Cross-validation is a general method of statistical analysis with various purposes, but mainly for model validation. It traditionally involves multiple rounds with three steps: the bi-partitioning of the data set, analysis of one partition, and validation/testing on the other one. Results from these rounds may be analyzed after some aggregation, or individually.
For a general introduction to the topic, see the survey of \citet{arlotSurveyCrossvalidationProcedures2010}.
There are multiple variants with tri-partitioning, to generate distinct validation and testing sets.
CV is most popular in predictive analytics to assess the accuracy of prediction on new (test) data, independent of the already-seen (training) data. The aggregation of the results yields a more accurate estimate of the model prediction error. It is also applied in model parameter selection, \eg, by \citet{varmaBiasErrorEstimation2006}, as an optimizer for classification parameters in SVM and shrunken centroid models. However, many statistical tools require modifications of the regular CV tools to adapt to different applications. For example, 
\citet{broCrossvalidationComponentModels2008} tested and reviewed the available CV techniques for Principal Component Analysis (PCA) only.
Domains with dependent data sets also require a revision of the CV methods to address dependency issues.
\citet{hijmansCrossvalidationSpeciesDistribution2012} addresses spatial sorting bias (a usual problem in species distribution models) using pairwise distance sampling.
\Eg, \citet{robertsCrossvalidationStrategiesData2017} examine various techniques and recommend that block CV should be used wherever dependence structures exist in a data set. Various problems arise during the parameter selection of the CV techniques themselves, \eg, \citet{fushikiEstimationPredictionError2011} examines bias correction of the $k$-fold CV in the prediction setup for small $k$-s with large samples.
Bayesian modifications are also studied in CV, \eg, by \citet{vehtariPracticalBayesianModel2017}.
\citet{isakssonCrossvalidationBootstrappingAre2008} state that CV and bootstrap are, in fact, unreliable in small samples, and promote Bayesian alternatives because the uncertainty introduced by the CV is too large. 
Different repetition techniques were also introduced on top of the CV methods to assess different problems, \eg,
for parameter tuning in classification and regression problems by \citet{krstajicCrossvalidationPitfallsWhen2014},
for optimizing the complexity of regression models for small data sets by \citet{filzmoserRepeatedDoubleCross2009}, and
for reducing the variability of the estimator by \citet{kimEstimatingClassificationError2009}.
Domain-specific problems also require the study of the CV techniques, \eg, \citet{wengerAssessingTransferabilityEcological2012} demonstrate the importance of considering model transferability in ecology based on different cross-validation approaches.

\paragraph{Uncertainty of rankings}
There are many approaches to measure the uncertainty among many rankings, but the most direct way to do so is to calculate the standard deviation of the rankings as if they were simply real vectors. For rankings $\pi_1,\dots, \pi_m$ of $n$ elements, it is
$
\operatorname{SD}=\sqrt{
\frac{1}{m}\sum_{i=1}^m\brp{\pi_i-\overline{\pi}}^2},
$
where $\overline{\pi}$ is simply the mean of the sample (or objects) rankings as vectors. This $L_2$ approach has been applied in many papers, \eg\ in refs.~\citep{faliveneInterpolationAlgorithmRanking2010,palmerComparisonSpatialPrediction2009,triantafilisFiveGeostatisticalModels2001,farshadfarVitroApplicationIntegrated2016}.
Aside from the above metric, \citet{barlowPredictingAssessingDominance1976} measure uncertainty with another metric taken from information theory.
\citet{rosanderStandardErrorMean1936} calculates the standard error of the mean ranking based on the rank-order correlation.
In refs.~\citep{lockwoodUncertaintyRankEstimation2002,zampetakisQuantifyingUncertaintyRanking2010},
the uncertainty of rankings is considered using Bayesian modeling of the raw data, from where the rankings originate.
\citet{zukRankingUncertainty2007} tackle the uncertainty in a ranking by introducing noise to the raw data, and comparing the original ranking to the noisy one with Top-K-List overlap and Kendall's Tau measure.

Aggregating rankings to obtain an optimal one is a problem naturally arising in many fields of science, from multi-criteria decision-making and information fusion to social choice.
In the previously cited work of~\citet{lockwoodUncertaintyRankEstimation2002}, the authors also dealt with the problem of finding an optimal ranking in their Bayesian framework.
The approach of \citet{pachecoRankingClassificationAlgorithms2018} uses the mean and standard deviation of the raw data to determine a final ranking.
\citet{tavanaeiUnsupervisedLearningRank2018} take into account the uncertainty of ranks by the weights of a parameterized rank aggregation process.
\citet{klementievUnsupervisedLearningAlgorithm2007} work with the rankings only, instead of using the properties of raw data, based on the principle of rewarding ranking agreements.
\citet{tehraniPreferenceLearningUsing2012} use the Choquet integral as an underlying model for representing ranking before the aggregation.
\citet{volkovsNewLearningMethods2014} introduces probabilistic models (building on a multinomial
generative process) for preference aggregation in unsupervised and supervised setups too.
There are countless other techniques for ranking aggregation, with a lot of them coming from \eg\ voting theory where preference and judgment aggregation are age-old problems \citep{Endriss2018,Mongin2012}.

The aforementioned references usually deal with the uncertainty of a set of alternative rankings, or see the uncertainty in the raw data itself. A distinctive contribution of this paper is that we grasp the uncertainty of a single ranking by looking at its partial rankings coming from manyfold CV.

\section{Methodology}

SRD requires the input data to be arranged in a matrix
form: rows ($1,\dots, n$) represent objects (statistical cases, compounds, features, \emph{etc.}, depending on the use case), whereas columns ($1,\dots, m$) represent the variables (solutions, models, methods, \emph{etc.}) to be compared.
There is one designated column containing a reference value for each object. These can be a previously established gold standard, an estimation, or even an aggregation from the input data (this last technique is also called \emph{data fusion}). The input matrix is transformed into a ranking matrix by ranking the values in each column from the smallest to the largest element. The resulting $n\times (m+1)$ matrix contains $m$ rankings $\pi_1,\dots, \pi_m$ associated to the variables and a reference ranking $\pi_r$.

\subsection{The Metric SRD is Built Upon}

The distance metric applied in the SRD framework is simply the $L_1$  norm or city block (Manhattan) distance of the rankings; it is Spearman's footrule if no ties are present in the ranking.
For rankings $\pi_i, \pi_j$, we will denote it with $d(\pi_i, \pi_j):=\sum_{k=1}^n \abs{\pi_{i}(k)-\pi_{j}(k)}$. In particular, the distance of $\pi_i$ from the reference will be denoted as $\SRD(\pi_i):=d(\pi_i, \pi_r)$.


Properties of the Spearman's footrule have been intensively studied over the symmetric group $S_n$, which contains the permutations over $1,\dots, n$.
The maximal distance is easy to compute:
\begin{equation}\label{eq:max}
M:=\underset{i,j}{\max}\ d(\pi_i, \pi_j) =
\begin{cases}
\frac{n^2}{2} &\text{if $n$ is even}\\
\frac{n^2-1}{2} &\text{if $n$ is odd}
\end{cases}
\end{equation}
The distance is right-invariant to the composition operation, meaning
$$d(\pi_i, \pi_j) = d(\pi_i\sigma, \pi_j\sigma)
\quad \forall \sigma \in S_n.$$
For convenience and interpretability, we usually normalize the distance by the maximal distance of Eq.~\ref{eq:max}:
\[
\SRDnorm(\pi):=\SRD(\pi)/M.
\]
The normalized $\SRDnorm$
values also make comparison possible if the
numbers of objects are different.

\citet{Diaconis_Graham_1977} showed that the distance is asymptotically normal (as $n\to \infty$) if we choose two permutations uniformly from $S_n$:
\[
d(\cdot, \cdot) \sim  \norm\brp{\frac{1}{3}n^2 + \bigO(n),\ \sqrt{\frac{2}{45}n^3+\bigO(n^2)}}.
\]
Due to right-invariance, the distribution is the same if we fix one of the permutations, as it happens in the SRD framework where we have a fixed reference ranking.
After normalization:
\begin{equation}\label{eq:srddist}
   \SRDnorm(\cdot) \sim  \norm\brp{\frac{2}{3},\ \sqrt{\frac{8}{45n}}}.
\end{equation}

Based on the above properties, a hypothesis test can be created to answer the question: Is a specific model ranking ($\pi_{model}$) close enough to the reference ranking ($\pi_r$)? The null hypothesis $H_0$ is that the ranking is uniformly selected from $S_n$. The test statistic is $\SRDnorm(\pi_{model})$, which is asymptotically normal for large $n$ under $H_0$ by Eq.~\ref{eq:srddist}. If ties are present or $n$ is small, the exact discrete distribution should be used~\citep{hebergerSumRankingDifferences2011}.


\subsection{Uncertainty of SRD}

As already mentioned, SRD values are deterministic by nature, but with CV, we can introduce uncertainty to them in order to compare two rankings with each other.

\subsubsection{Wilcoxon Signed-Rank Test}

The Wilcoxon signed-rank test was proposed for CV purposes in ref.~\citep{KOLLARHUNEK2013139}.
Take random subsets ($A_1,\dots,A_k$) of size $n-\ceil{n/k}$ from the $n$ rows/objects. Let $\SRDnorm_{j,i}$ denote the $\SRDnorm$ (from the reference) on fold $A_i$ for the ranking $\pi_j$. This results in a paired sample of size $k$ for the two models to compare:
\[
\brc{
\brp{\SRDnorm_{1,1}, \SRDnorm_{2,1}},
\brp{\SRDnorm_{1,2}, \SRDnorm_{2,2}},
\dots,
\brp{\SRDnorm_{1,k}, \SRDnorm_{2,k}}
}.
\]
To this, we apply the signed-rank test. We take the sub-sample (of size $k_r<k$) containing the non-zero absolute differences $|\SRDnorm_{1,i} -\SRDnorm_{2,i}|$, and then rank this subsample. The test-statistic $W$ comes from
\[
W^+ = \sum_{i=1}^k
\I\brp{\SRDnorm_{1,i}>\SRDnorm_{2,i}}
\cdot R_i,
\quad\text{and}\quad
W^- = \sum_{i=1}^k
\I\brp{\SRDnorm_{1,i}<\SRDnorm_{2,i}}
\cdot R_i,
\]
where $\I()$ is the indicator function, and $R_i$ is the rank of the $i$th fold in the aforementioned subsample. $W$ can be $\min(W^+, W^-)$, $W^+-W^-$, or $W^+$ itself. In each case, $W$ has a specific distribution (depending on $k_r$) under the $H_0$ that the difference between the pairs follows a symmetric distribution around zero.


\subsubsection{Dietterich \texorpdfstring{$5\times2$}{5x2} CV \texorpdfstring{$\td$}{t}-test}\label{sec:diett}

\citet{dietterichApproximateStatisticalTests1998} proposed this test for determining whether there is a significant difference between the error rates of the two classifiers. Here, we show how to apply the test within the SRD framework for comparing the rankings of two models ($\pi_1, \pi_2$).

We start by taking random subsets ($A_1,\dots,A_k$) and their complements ($A_1^c,\dots,A_k^c$)
from the $n$ rows (objects) of sizes $\floor*{\frac{n}2}$ and $\ceil*{\frac{n}2}$. In the original paper and in most applications, $k=5$.
Let $\SRDnorm_{j,i}$ denote once again the $\SRDnorm$ (from the reference) on fold $A_i$ for the ranking $\pi_j$, and let $\SRDnorm_{j,i}^c$ denote the same on the complement $A_i^c$.

First, look only at the fold $(A_i, A_i^c)$.
Calculate the differences of the normalized SRDs in both subsets,
\[
\Delta_{i} := \SRDnorm_{1,i} - \SRDnorm_{2,i}
\qquad\text{and}\qquad
\Delta_{i^c} := \SRDnorm_{1,i}^c - \SRDnorm_{2,i}^c,
\]
and then calculate their average and sample variance:
\[
\overline{\Delta_i}= \frac12\brp{\Delta_i +\Delta_{i^c}},
\qquad
s_{\Delta,i}^2 =
\brp{\Delta_i-\overline{\Delta_i}}^2
+
\brp{\Delta_{i^c}-\overline{\Delta_i}}^2 .
\]
For large enough $n$, $\Delta_{i}$ and $\Delta_{i^c}$ are asymptotically normal under the $H_0$ of uniformly selecting a ranking from $S_n$, thus
we can get an approximately $\td$-distributed test statistic:
\[
\frac{\Delta_{i}}
{
\sqrt{\frac1k
\sum_{i=1}^k
s_{\Delta,i}^2
}}\sim \td_{k},
\]
and the same applies to the complementing fold.
This way, we have $2k$ different test statistics, of which we can choose any for the evaluation of the hypothesis test.

\subsubsection{Alpaydin \texorpdfstring{$5\times2$}{5x2} CV \texorpdfstring{$\F$}{F}-test}

The $5\times2$ CV $\F$-test of \citet{alpaydinCombinedCvTest1999}
was proposed as an improvement on the CV $\td$-test of \citet{dietterichApproximateStatisticalTests1998}. Here, we show how to apply the test within the SRD framework for comparing the rankings of two models ($\pi_1, \pi_2$).
Up until the calculation of the
$\Delta_{i}$, $\Delta_{i^c}$ pairs and their sample variances for each fold, the setup is the same as in Section~\ref{sec:diett}. However, the test statistic is approximately $\F$-distributed:
\[
\frac{
\frac1{2}
\brp{\Delta_i^2 +\Delta_{i^c}^2}}
{
s_{\Delta,i}^2
}\sim \F_{2,1}.
\]
Aggregating these for all folds results in:
\[
\frac{
\frac1{2k}
\sum_{i=1}^k
\brp{\Delta_{i}^2 +\Delta_{i^c}^2}}
{
\frac1k
\sum_{i=1}^k
s_{\Delta,i}^2
}\sim \F_{2k,k}.
\]
Note that this assumes the independence of the repeated measurements on the folds, and again, the independence of the numerator and denominator too.
Instead of aggregating the statistics of the folds, one can aggregate the dependent $p$-values coming from the folds separately.
The correct aggregation of $p$-values would only assume the independence of the numerator and denominator.

\section{Evaluation}

How can the best fitting option be selected in a complex decision situation where all the competing solutions seem fair in some way? A formal approach to this, commonly used in economics, is axiomatization. We characterize the methods by the properties they satisfy (Pareto optimality, symmetry, monotonicity, \emph{etc.}). We clarify which properties are important for us and then choose the method that best suits our needs.

Statistical tests do not fit into this scheme readily. Their behavior is stochastic rather than binary, meaning, none of them satisfy properties 100$\%$ of the time. Nevertheless, when choosing from tests, an axiomatic mindset can be useful; the idea is to come up with scenarios that the tests will likely face in practice, and then observe through a simulation which test is more apt in which scenario. This type of analysis is especially suitable for rankings, as the differences between two rankings can be characterized fairly well.

We have evaluated three hybrid tests (Wilcoxon, Dietterich, and Alpaydin) under various parametrizations. The aim was to find the one that is the most efficacious in categorizing solutions. We have used the number of folds as parameters and varied it between 5 and 10.

\subsection{Scenarios}\label{sec:scenarios}

We analyzed nine scenarios under three assumptions on the size of the rankings ($n=7$, $n=13$, and $n=32$). The size options aim to represent the typical data sizes. There is no point in further increasing $n$ as anything that is true for $n=32$ will probably stay true for larger row sizes. Moreover, note that $n=7$ is an odd number and almost too small for CV. However, since practitioners will not refrain from applying cross-validation for suboptimal data sizes, we felt the need to test this case as well. For $n=7$ and $k\ge7$, the Wilcoxon reduces to leave-one-out CV. In particular, if the number of folds, $k$, exceeds seven, we are forced to use bootstrapping; some rows are left out more than once.

All rankings that come from the same distribution are alike, rankings that come from different distributions are different in their own way\footnote{We are slightly paraphrasing Lev Tolstoy's Anna Karenina here.}—hence we have looked at six scenarios for type II errors but only three for type I errors.
These scenarios cover all typical situations, and other scenarios are unlikely to yield a new aspect as they would constitute a transition in between these.

\begin{figure}[!ht]
    \centering
    \includegraphics{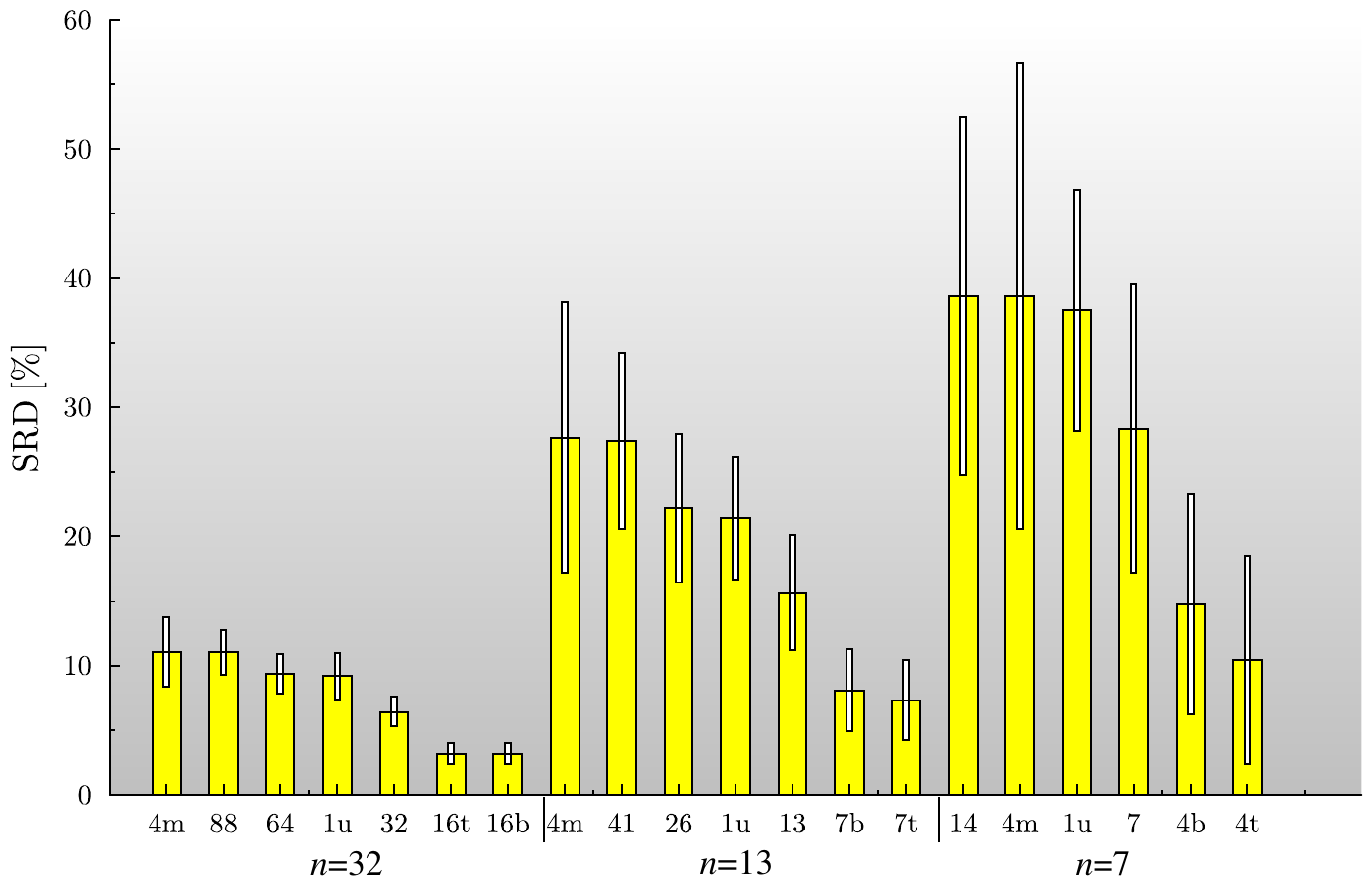}
    \caption{Transformed rankings average distance from the reference $\pm$ standard deviation in normalized SRD}
    \label{fig:trans_SRD}
\end{figure}

In the following list, we have described the scenarios in detail. We started with a reference ranking, which is just the ordered list of numbers from 1 to $n$. Then, we compose two additional rankings, denoted by A and B, by making some transformations to the reference ranking. For checking type I error, we drew both rankings from the same distribution (Scenarios 1-3). In type II scenarios, the rankings were constructed in different ways (Scenarios 4-9). Table~\ref{tab:rank_trans} displays the ranking transformations that were used to produce the rankings, and Figure~\ref{fig:trans_SRD} shows their average distance from the reference.

\begin{table}\caption{Ranking transformations used in the scenarios}\label{tab:rank_trans}
\begin{tabular}{ll}
Identifier & Description  \\ \hline
\multirow{2}{*}{$x$}          & \multirow{2}{14cm}{We apply $x$ number of random inversions (switching of neighboring elements) on the reference ranking.}                                                                                                                           \\ \\
\multirow{2}{*}{$x$t}       & \multirow{2}{14cm}{We apply $x$ number of random inversions on the top $\floor*{n/2}$ positions of the reference ranking.}                                                                                                                      \\ \\
\multirow{2}{*}{$x$b}         & \multirow{2}{14cm}{We apply $x$ number of random inversions on the bottom $\ceil*{n/2}$ positions of the reference ranking.}                                                                                                                     \\ \\
\multirow{2}{*}{1u}         & \multirow{2}{14cm}{We select a random element (the underdog) from the bottom $\ceil*{n/2}$ positions of the reference ranking and switch it with the first element.}                                                                                          \\ \\
\multirow{3}{*}{4m}         &\multirow{3}{14cm}{We consecutively select four random positions between 1 and $n-\ceil*{n/4}$ from the reference ranking and switch the selected element at position $s$ with the element at position $s+\ceil*{n/4}$.} \\ \\
\end{tabular}
\end{table}

A scenario is defined as a pair $(a|b)$, where $a$ and $b$ refer to the transformations used to create rankings A and B respectively. For the definition of the transformations see Table~\ref{tab:rank_trans}.

\begin{enumerate}
\item $(2n|2n)$: This scenario investigates what happens if both rankings are drawn from the same distribution and their distance to the reference ranking is relatively large.

\item $(n|n)$: Similar to the previous one but with fewer inversions, so the rankings are closer to the reference.

\item RT I.: We picked a transformation uniformly and randomly from the set $2n,\ n,\ (n/2)t$, $(n/2)b,\ 1u,\ 4m$. Both rankings A and B were drawn from the selected distribution. This scenario demonstrates what can we expect from a CV method regarding type I error when we do not have prior information on the rankings' distribution.

\item $(2n|n/2)$: This scenario investigates what happens if the rankings are drawn from different distributions, that is, if their mean distance to the reference ranking is different, because the second ranking is a result of fewer inversions, so it is closer to the reference.

\item $(n|n/2)$: A noisier variant of the previous scenario. Here the average distance between rankings is smaller, hence, it is more difficult to distinguish between the rankings.

\item $((n/2)t|(n/2)b)$: This scenario shows what happens when the data is structured. For instance, for data collected from two periods, the solutions perform differently on the first and second periods. Note that the expected distance from the reference is approximately the same for both A and B (\emph{cf.}~Figure~\ref{fig:trans_SRD}).

\item $(2n|1u)$: This scenario tests the presence of outliers. Rankings A and B are of the same distance from the reference. However, the former is constructed by applying many small inversions, while in ranking B, we only swap one pair of elements. To illustrate this scenario, let us borrow an example from sports. Ranking A shows how the actual result of a sporting event, \eg\ the soccer World Cup, differs from the preliminary ranking. Ranking B is the same as the preliminary ranking except that Burkina Faso wins, relegating Brazil to the second half of the points table. The preferability of ranking A or B depends on the application, but a CV method should be able to distinguish between the two rankings.

\item $(x|4m)$ Similar to the previous scenario, but having more, albeit less extreme, outliers. The number of inversions for ranking A, $x$, is chosen in such a way that the expected distance from the reference for both rankings is approximately the same.

\item RT II.: We picked a transformation uniformly and randomly from the set $2n,\ n,\ (n/2)t$, $(n/2)b,\ 1u,\ 4m$. Ranking A was drawn from the selected distribution, and for B, we picked another transformation randomly. This scenario demonstrates what can we expect from a CV method regarding type II error when we do not have prior information on the rankings' distribution.
\end{enumerate}

\begin{figure}[!ht]
    \centering
    \includegraphics{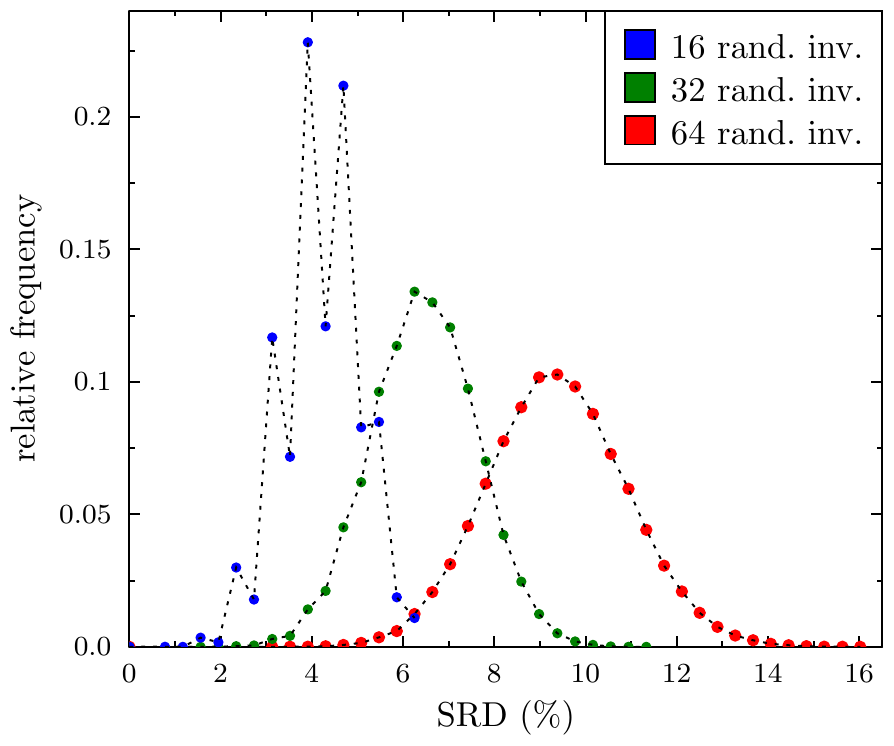}
    \caption{The discrete distribution of the distances from reference under different transformations in the $n=32$ case}
    \label{fig:SRD_64_32_16}
\end{figure}

Figure~\ref{fig:SRD_64_32_16} shows the relative frequency of distances between a perturbed ranking and the reference ranking in the $n=32$ case. As the number of inversions grows the discrete distribution of the distance values approximates the normal distribution.

Manhattan distance satisfies the triangle inequality, hence if one ranking is close to the reference and the other is not, then they fall far from each other. There is significant overlap among the distributions generated by the distances of 16 and 32 random inversions. Thus we cannot expect any CV method to tell them apart 100$\%$ of the time. Note that just because two rankings are of the same distance from the reference it does not necessarily mean that they are close to each other. Only, in such cases we cannot eliminate the possibility, so it is difficult to quantify the ideal rejection rate.

In the 64|16 scenario, however, the distributions overlap only at their tails. There is a very low probability that the rankings are of a similar distance from the reference, and even if they are, they may differ in completely disparate segments. CV methods should be able to distinguish between such rankings in the vast majority of the cases.

\subsection{Simulation Results}\label{sec:sim_results}

For each scenario and data size, we performed 10 simulation rounds, each comprising 100~000 runs. We randomly generated instances of rankings A and B, applied the CV technique, and noted whether it accepted or rejected the null hypothesis at $5\%$ level of significance. Tables~\ref{tab:results_n32}-\ref{tab:results_n7} show the average rate of rejections in the 10 rounds. In type I scenarios (1–3) the lower the rejection rate the better—the rankings come from the same distribution, so the CV method is expected to accept the null hypothesis.\footnote{Note that, if all the rigorous assumptions of the applied statistical tests held firm, these numbers would be around $5\%$. Both \citet{dietterichApproximateStatisticalTests1998} and \citet{alpaydinCombinedCvTest1999} knew well that their assumptions are not satisfied in general but dismissed concerns based on the empirical success of the tests.} In type II scenarios (4–9), the opposite is true; the higher the rejection rate, the better. We also observed the standard error on the average rejection rate of the 10 rounds, which falls under 0.002 in all scenarios.

There are several conclusions that we can infer from Tables~\ref{tab:results_n32}–\ref{tab:results_n7}:

\begin{itemize}
    \item No CV method excels in both type I and II situations. 
    \item The Dietterich and Alpaydin tests rarely reject the null hypothesis. Thus, they excel in type I scenarios but perform poorly in type II scenarios.

    \item The opposite is true for Wilcoxon: it shines in type II situations but in type I cases the error rate is nowhere near the theoretical 5\%.
    \item With some rare exceptions, increasing the number of folds raises the rate of rejection for all methods. This results in a trade-off between the efficiency of methods in type I and type II scenarios.
\end{itemize}

As the tables show, different techniques excel in different scenarios. If practitioners have some preliminary knowledge about the distribution of the rankings that the investigated methods produce, they can choose the appropriate CV method. In many cases, however, these distributions are unknown. Therefore, to select the most suitable method for generic purposes, we need to dig deeper into the data.

\begin{table}[!ht]
\scriptsize
\caption{Null hypothesis testing for $n=32$ -- rejection rate ($\%$) in different scenarios}\label{tab:results_n32}
\begin{tabular}{l|lll|llllll}
\multirow{2}{2.5 cm}{CV method / number of folds}                                              & \multicolumn{3}{c|}{type I. scenarios}                                                     & \multicolumn{6}{c}{type II. scenarios}                                                                                                                                                  \\
 & 64|64                        & 32/32                        & RT I                         & 64|16                        & 32|16                        & 16t|16b                      & 64|1u                        & 88|4m                        & RT II                        \\ \hline
Wilcoxon 5                                                         & \cellcolor[HTML]{FBCCCE}35.5 & \cellcolor[HTML]{FBD3D6}30.3 & \cellcolor[HTML]{FBC8CB}37.5 & \cellcolor[HTML]{6CC283}95.1 & \cellcolor[HTML]{A9DBB8}61.1 & \cellcolor[HTML]{F8FBFC}16.8 & \cellcolor[HTML]{BAE2C6}51.5 & \cellcolor[HTML]{B8E1C4}53.0 & \cellcolor[HTML]{92D1A3}74.1 \\
Wilcoxon 6                                                         & \cellcolor[HTML]{FBCBCE}35.7 & \cellcolor[HTML]{FBD3D6}30.3 & \cellcolor[HTML]{FBC8CB}37.8 & \cellcolor[HTML]{6BC182}95.9 & \cellcolor[HTML]{A6DAB5}62.6 & \cellcolor[HTML]{FBFCFE}15.6 & \cellcolor[HTML]{BAE1C6}51.8 & \cellcolor[HTML]{B8E1C4}53.0 & \cellcolor[HTML]{90D0A2}75.2 \\
Wilcoxon 7                                                         & \cellcolor[HTML]{FBBFC2}43.7 & \cellcolor[HTML]{FBC8CB}37.5 & \cellcolor[HTML]{FBC0C3}42.9 & \cellcolor[HTML]{68C080}97.3 & \cellcolor[HTML]{9BD5AB}69.1 & \cellcolor[HTML]{F0F7F5}21.5 & \cellcolor[HTML]{B0DEBE}57.0 & \cellcolor[HTML]{A6D9B5}62.8 & \cellcolor[HTML]{85CC99}81.2 \\
Wilcoxon 8                                                         & \cellcolor[HTML]{FBBABC}47.2 & \cellcolor[HTML]{FBC3C6}40.8 & \cellcolor[HTML]{FBBCBE}45.9 & \cellcolor[HTML]{67C07F}97.9 & \cellcolor[HTML]{95D3A6}72.2 & \cellcolor[HTML]{ECF6F1}24.0 & \cellcolor[HTML]{AADBB9}60.4 & \cellcolor[HTML]{A1D7B0}65.8 & \cellcolor[HTML]{82CB96}83.1 \\
Wilcoxon 9                                                         & \cellcolor[HTML]{FAADAF}55.9 & \cellcolor[HTML]{FBB5B8}50.3 & \cellcolor[HTML]{FAB1B3}53.2 & \cellcolor[HTML]{65BF7D}98.9 & \cellcolor[HTML]{88CD9B}79.7 & \cellcolor[HTML]{DDF0E4}32.3 & \cellcolor[HTML]{9FD7AF}66.5 & \cellcolor[HTML]{94D2A5}73.0 & \cellcolor[HTML]{7CC891}86.3 \\
Wilcoxon 10                                                        & \cellcolor[HTML]{FAA7AA}59.4 & \cellcolor[HTML]{FAB0B3}53.5 & \cellcolor[HTML]{FAAEB1}54.9 & \cellcolor[HTML]{65BF7D}99.1 & \cellcolor[HTML]{85CC98}81.2 & \cellcolor[HTML]{D5EDDE}36.5 & \cellcolor[HTML]{9DD6AD}67.8 & \cellcolor[HTML]{8ED0A0}76.4 & \cellcolor[HTML]{7AC88F}87.3 \\
Dietterich 5                                                       & \cellcolor[HTML]{ECF5F2}3.0  & \cellcolor[HTML]{DFF0E6}2.7  & \cellcolor[HTML]{DAEEE2}2.6  & \cellcolor[HTML]{EEF7F3}22.5 & \cellcolor[HTML]{F9AEB0}6.8  & \cellcolor[HTML]{F86F71}0.6  & \cellcolor[HTML]{F87D7F}2.0  & \cellcolor[HTML]{F88689}3.0  & \cellcolor[HTML]{FBDEE1}11.6 \\
Dietterich 6                                                       & \cellcolor[HTML]{F7FAFB}3.2  & \cellcolor[HTML]{EDF5F2}3.0  & \cellcolor[HTML]{E7F3ED}2.8  & \cellcolor[HTML]{EBF5F0}24.3 & \cellcolor[HTML]{FAB4B7}7.5  & \cellcolor[HTML]{F86F71}0.7  & \cellcolor[HTML]{F87779}1.5  & \cellcolor[HTML]{F8898B}3.2  & \cellcolor[HTML]{FBE5E8}12.3 \\
Dietterich 7                                                       & \cellcolor[HTML]{FCFCFF}3.3  & \cellcolor[HTML]{EFF6F4}3.0  & \cellcolor[HTML]{EFF7F4}3.0  & \cellcolor[HTML]{E9F5EF}25.2 & \cellcolor[HTML]{FAB7BA}7.7  & \cellcolor[HTML]{F87072}0.7  & \cellcolor[HTML]{F87476}1.2  & \cellcolor[HTML]{F88A8C}3.3  & \cellcolor[HTML]{FBE8EB}12.6 \\
Dietterich 8                                                       & \cellcolor[HTML]{FCFCFF}3.5  & \cellcolor[HTML]{FAFBFD}3.2  & \cellcolor[HTML]{F9FBFC}3.2  & \cellcolor[HTML]{E8F4ED}26.2 & \cellcolor[HTML]{FABBBE}8.2  & \cellcolor[HTML]{F87072}0.8  & \cellcolor[HTML]{F87274}1.0  & \cellcolor[HTML]{F88C8E}3.5  & \cellcolor[HTML]{FBECEF}13.0 \\
Dietterich 9                                                       & \cellcolor[HTML]{FCFCFF}3.6  & \cellcolor[HTML]{FBFBFE}3.3  & \cellcolor[HTML]{FCFCFF}3.4  & \cellcolor[HTML]{E7F4ED}26.7 & \cellcolor[HTML]{FABDBF}8.3  & \cellcolor[HTML]{F87173}0.8  & \cellcolor[HTML]{F87173}0.8  & \cellcolor[HTML]{F98E90}3.7  & \cellcolor[HTML]{FBEEF1}13.2 \\
Dietterich 10                                                      & \cellcolor[HTML]{FCFCFF}3.8  & \cellcolor[HTML]{FCFCFF}3.4  & \cellcolor[HTML]{FCFCFF}3.5  & \cellcolor[HTML]{E6F3EC}27.2 & \cellcolor[HTML]{FABFC1}8.5  & \cellcolor[HTML]{F87173}0.9  & \cellcolor[HTML]{F87072}0.7  & \cellcolor[HTML]{F98F91}3.8  & \cellcolor[HTML]{FBF1F4}13.5 \\
Alpaydin 5                                                         & \cellcolor[HTML]{D5ECDD}2.4  & \cellcolor[HTML]{C2E4CD}2.1  & \cellcolor[HTML]{B6DFC3}1.8  & \cellcolor[HTML]{DFF1E6}30.8 & \cellcolor[HTML]{FAB2B5}7.3  & \cellcolor[HTML]{F86C6E}0.3  & \cellcolor[HTML]{F88183}2.4  & \cellcolor[HTML]{F88486}2.7  & \cellcolor[HTML]{FBFCFE}15.5 \\
Alpaydin 6                                                         & \cellcolor[HTML]{DDEFE5}2.6  & \cellcolor[HTML]{C6E6D0}2.1  & \cellcolor[HTML]{BCE2C8}1.9  & \cellcolor[HTML]{D5ECDD}36.8 & \cellcolor[HTML]{FABEC1}8.5  & \cellcolor[HTML]{F86B6D}0.3  & \cellcolor[HTML]{F87E80}2.1  & \cellcolor[HTML]{F88587}2.8  & \cellcolor[HTML]{F6FAFA}18.1 \\
Alpaydin 7                                                         & \cellcolor[HTML]{E4F2EB}2.8  & \cellcolor[HTML]{CBE8D5}2.2  & \cellcolor[HTML]{BFE3CB}2.0  & \cellcolor[HTML]{CBE8D5}42.3 & \cellcolor[HTML]{FAC9CC}9.5  & \cellcolor[HTML]{F86B6D}0.2  & \cellcolor[HTML]{F87C7E}1.9  & \cellcolor[HTML]{F88689}2.9  & \cellcolor[HTML]{F2F8F6}20.6 \\
Alpaydin 8                                                         & \cellcolor[HTML]{F1F7F5}3.0  & \cellcolor[HTML]{D1EADA}2.4  & \cellcolor[HTML]{C7E6D1}2.2  & \cellcolor[HTML]{C2E5CD}47.0 & \cellcolor[HTML]{FAD5D8}10.7 & \cellcolor[HTML]{F86A6C}0.2  & \cellcolor[HTML]{F87B7D}1.8  & \cellcolor[HTML]{F8888B}3.1  & \cellcolor[HTML]{EEF6F3}22.9 \\
Alpaydin 9                                                         & \cellcolor[HTML]{FCFCFF}3.3  & \cellcolor[HTML]{D8EDE0}2.5  & \cellcolor[HTML]{CFE9D8}2.3  & \cellcolor[HTML]{BBE2C7}51.3 & \cellcolor[HTML]{FBE1E4}11.9 & \cellcolor[HTML]{F86A6C}0.2  & \cellcolor[HTML]{F87A7C}1.7  & \cellcolor[HTML]{F88A8C}3.3  & \cellcolor[HTML]{EAF5EF}25.0 \\
Alpaydin 10                                                        & \cellcolor[HTML]{FCFCFF}3.5  & \cellcolor[HTML]{E1F1E8}2.7  & \cellcolor[HTML]{D6ECDE}2.5  & \cellcolor[HTML]{B4DFC1}54.9 & \cellcolor[HTML]{FBEDF0}13.0 & \cellcolor[HTML]{F86A6C}0.2  & \cellcolor[HTML]{F87A7C}1.7  & \cellcolor[HTML]{F88C8E}3.5  & \cellcolor[HTML]{E6F4EC}26.8
\end{tabular}
\caption{Null hypothesis testing for $n=13$ -- rejection rate ($\%$) in different scenarios}\label{tab:results_n13}
\begin{tabular}{l|lll|llllll}
\multirow{2}{2.5 cm}{CV method / number of folds}                                                 & \multicolumn{3}{c|}{type I. scenarios}                                                     & \multicolumn{6}{c}{type II. scenarios}                                                                                                                                                  \\
 & \multicolumn{1}{r}{26|26}    & \multicolumn{1}{r}{13|13}    & \multicolumn{1}{r|}{RT I}    & \multicolumn{1}{r}{26|7}     & \multicolumn{1}{r}{13|7}     & \multicolumn{1}{r}{7t|7b}    & \multicolumn{1}{r}{26|1u}    & \multicolumn{1}{r}{41|4m}    & \multicolumn{1}{r}{RT II}    \\ \hline
Wilcoxon 5                                                           & \cellcolor[HTML]{FAA5A8}30.4 & \cellcolor[HTML]{FBB6B9}25.3 & \cellcolor[HTML]{FA9B9E}33.4 & \cellcolor[HTML]{85CC98}64.6 & \cellcolor[HTML]{C5E6D0}33.7 & \cellcolor[HTML]{F0F8F5}13.1 & \cellcolor[HTML]{B9E1C5}39.7 & \cellcolor[HTML]{AEDDBC}45.0 & \cellcolor[HTML]{8FD0A1}59.9 \\
Wilcoxon 6                                                           & \cellcolor[HTML]{FAB1B4}26.9 & \cellcolor[HTML]{FBC2C5}21.8 & \cellcolor[HTML]{FAA6A8}30.3 & \cellcolor[HTML]{8BCF9E}61.7 & \cellcolor[HTML]{CDE9D6}30.1 & \cellcolor[HTML]{F7FAFA}10.1 & \cellcolor[HTML]{C1E4CC}36.0 & \cellcolor[HTML]{B6E0C2}41.3 & \cellcolor[HTML]{95D2A6}57.0 \\
Wilcoxon 7                                                           & \cellcolor[HTML]{F98E90}37.7 & \cellcolor[HTML]{FAA2A4}31.5 & \cellcolor[HTML]{F98789}39.6 & \cellcolor[HTML]{75C58A}72.5 & \cellcolor[HTML]{B7E0C3}40.8 & \cellcolor[HTML]{E8F4EE}16.9 & \cellcolor[HTML]{ADDCBB}45.3 & \cellcolor[HTML]{98D4A8}55.7 & \cellcolor[HTML]{7EC992}68.0 \\
Wilcoxon 8                                                           & \cellcolor[HTML]{FA9294}36.3 & \cellcolor[HTML]{FAA7A9}30.0 & \cellcolor[HTML]{F98C8E}38.1 & \cellcolor[HTML]{77C68C}71.4 & \cellcolor[HTML]{BAE1C6}39.4 & \cellcolor[HTML]{EBF5F0}15.7 & \cellcolor[HTML]{B0DEBE}43.9 & \cellcolor[HTML]{9AD5AB}54.4 & \cellcolor[HTML]{80CA94}67.2 \\
Wilcoxon 9                                                           & \cellcolor[HTML]{F97173}46.4 & \cellcolor[HTML]{F98486}40.5 & \cellcolor[HTML]{F96C6E}47.8 & \cellcolor[HTML]{65BF7D}79.8 & \cellcolor[HTML]{A3D8B2}50.3 & \cellcolor[HTML]{DAEEE2}23.9 & \cellcolor[HTML]{9FD7AF}52.2 & \cellcolor[HTML]{86CC99}64.2 & \cellcolor[HTML]{72C488}73.7 \\
Wilcoxon 10                                                          & \cellcolor[HTML]{F8696B}48.5 & \cellcolor[HTML]{F98082}41.7 & \cellcolor[HTML]{F96B6D}48.0 & \cellcolor[HTML]{63BE7B}80.6 & \cellcolor[HTML]{A1D7B0}51.2 & \cellcolor[HTML]{D5EDDE}26.0 & \cellcolor[HTML]{9ED6AE}52.8 & \cellcolor[HTML]{81CB95}66.4 & \cellcolor[HTML]{70C487}74.5 \\
Dietterich 5                                                         & \cellcolor[HTML]{EAF4EF}3.7  & \cellcolor[HTML]{E4F2EA}3.5  & \cellcolor[HTML]{D4EBDC}3.1  & \cellcolor[HTML]{F7FAFB}9.9  & \cellcolor[HTML]{FAC6C9}4.6  & \cellcolor[HTML]{F87678}0.7  & \cellcolor[HTML]{F99699}2.3  & \cellcolor[HTML]{FAC1C4}4.4  & \cellcolor[HTML]{FBFCFF}7.7  \\
Dietterich 6                                                         & \cellcolor[HTML]{F9FBFC}4.1  & \cellcolor[HTML]{F0F7F5}3.9  & \cellcolor[HTML]{E7F3ED}3.6  & \cellcolor[HTML]{F5FAF9}10.8 & \cellcolor[HTML]{FAD1D3}5.1  & \cellcolor[HTML]{F8797B}0.8  & \cellcolor[HTML]{F99496}2.1  & \cellcolor[HTML]{FAC9CC}4.8  & \cellcolor[HTML]{FAFCFE}8.3  \\
Dietterich 7                                                         & \cellcolor[HTML]{FCFCFF}4.5  & \cellcolor[HTML]{FCFCFF}4.2  & \cellcolor[HTML]{F5F9F9}4.0  & \cellcolor[HTML]{F4F9F8}11.4 & \cellcolor[HTML]{FBD8DB}5.5  & \cellcolor[HTML]{F87B7D}0.9  & \cellcolor[HTML]{F99395}2.1  & \cellcolor[HTML]{FACFD2}5.1  & \cellcolor[HTML]{F9FBFD}8.7  \\
Dietterich 8                                                         & \cellcolor[HTML]{FCFBFE}4.7  & \cellcolor[HTML]{FCFBFE}4.5  & \cellcolor[HTML]{FCFCFF}4.4  & \cellcolor[HTML]{F3F9F7}11.9 & \cellcolor[HTML]{FBDEE1}5.8  & \cellcolor[HTML]{F87D7F}1.0  & \cellcolor[HTML]{F99295}2.1  & \cellcolor[HTML]{FAD5D8}5.4  & \cellcolor[HTML]{F9FBFC}9.1  \\
Dietterich 9                                                         & \cellcolor[HTML]{FCFAFD}5.0  & \cellcolor[HTML]{FCFBFE}4.7  & \cellcolor[HTML]{FCFBFE}4.6  & \cellcolor[HTML]{F2F8F6}12.3 & \cellcolor[HTML]{FBE5E8}6.2  & \cellcolor[HTML]{F88082}1.2  & \cellcolor[HTML]{F99294}2.0  & \cellcolor[HTML]{FBDBDE}5.7  & \cellcolor[HTML]{F8FBFC}9.4  \\
Dietterich 10                                                        & \cellcolor[HTML]{FCF9FC}5.2  & \cellcolor[HTML]{FCFAFD}4.9  & \cellcolor[HTML]{FCFAFD}4.8  & \cellcolor[HTML]{F1F8F6}12.7 & \cellcolor[HTML]{FBE8EB}6.3  & \cellcolor[HTML]{F88285}1.3  & \cellcolor[HTML]{F99395}2.1  & \cellcolor[HTML]{FBDDE0}5.8  & \cellcolor[HTML]{F7FAFB}9.7  \\
Alpaydin 5                                                           & \cellcolor[HTML]{CAE7D3}2.8  & \cellcolor[HTML]{BFE3CB}2.5  & \cellcolor[HTML]{B3DEC0}2.2  & \cellcolor[HTML]{F5FAF9}10.7 & \cellcolor[HTML]{FAB5B8}3.8  & \cellcolor[HTML]{F86E70}0.3  & \cellcolor[HTML]{F99092}1.9  & \cellcolor[HTML]{FAC0C3}4.3  & \cellcolor[HTML]{FAFBFD}8.6  \\
Alpaydin 6                                                           & \cellcolor[HTML]{D6ECDE}3.2  & \cellcolor[HTML]{C6E6D0}2.7  & \cellcolor[HTML]{B9E0C5}2.3  & \cellcolor[HTML]{F1F8F6}12.7 & \cellcolor[HTML]{FABFC2}4.3  & \cellcolor[HTML]{F86D6F}0.2  & \cellcolor[HTML]{F98E90}1.8  & \cellcolor[HTML]{FACCCF}4.9  & \cellcolor[HTML]{F7FAFB}10.0 \\
Alpaydin 7                                                           & \cellcolor[HTML]{E1F1E7}3.4  & \cellcolor[HTML]{CEE9D7}2.9  & \cellcolor[HTML]{BFE3CA}2.5  & \cellcolor[HTML]{EDF6F2}14.5 & \cellcolor[HTML]{FAC9CB}4.7  & \cellcolor[HTML]{F86C6E}0.2  & \cellcolor[HTML]{F88D8F}1.8  & \cellcolor[HTML]{FAD6D9}5.4  & \cellcolor[HTML]{F4F9F8}11.4 \\
Alpaydin 8                                                           & \cellcolor[HTML]{EFF6F3}3.8  & \cellcolor[HTML]{D9EEE1}3.2  & \cellcolor[HTML]{C9E7D3}2.8  & \cellcolor[HTML]{E9F5EF}16.5 & \cellcolor[HTML]{FAD5D7}5.3  & \cellcolor[HTML]{F86B6D}0.1  & \cellcolor[HTML]{F98D90}1.8  & \cellcolor[HTML]{FBE1E3}5.9  & \cellcolor[HTML]{F1F8F6}12.7 \\
Alpaydin 9                                                           & \cellcolor[HTML]{FAFBFD}4.1  & \cellcolor[HTML]{E3F1E9}3.5  & \cellcolor[HTML]{D1EADA}3.0  & \cellcolor[HTML]{E6F3EC}18.2 & \cellcolor[HTML]{FBDFE2}5.8  & \cellcolor[HTML]{F86B6D}0.1  & \cellcolor[HTML]{F98E90}1.9  & \cellcolor[HTML]{FBE9EC}6.3  & \cellcolor[HTML]{EEF7F3}14.0 \\
Alpaydin 10                                                          & \cellcolor[HTML]{FCFBFE}4.5  & \cellcolor[HTML]{EFF6F3}3.8  & \cellcolor[HTML]{D9EEE1}3.2  & \cellcolor[HTML]{E2F2E9}19.9 & \cellcolor[HTML]{FBE9EC}6.3  & \cellcolor[HTML]{F86B6D}0.1  & \cellcolor[HTML]{F99092}1.9  & \cellcolor[HTML]{FBF2F5}6.8  & \cellcolor[HTML]{ECF6F1}15.3
\end{tabular}
\caption{Null hypothesis testing for $n=7$ -- rejection rate ($\%$) in different scenarios}\label{tab:results_n7}
\begin{tabular}{l|lll|llllll}
\multirow{2}{2.5 cm}{CV method / number of folds}                                                & \multicolumn{3}{c|}{type I. scenarios}                                                     & \multicolumn{6}{c}{type II. scenarios}                                                                                                                                                  \\
& 14|14                        & 7|7                          & RT I                         & 14|4                         & 7|4                          & 4t|4b                        & 14|1u                        & 14|4m                        & RT II                        \\ \hline
Wilcoxon 5                                                           & \cellcolor[HTML]{FBCCCE}19.5 & \cellcolor[HTML]{FCDBDE}15.4 & \cellcolor[HTML]{FBD1D4}18.1 & \cellcolor[HTML]{B7E1C4}31.2 & \cellcolor[HTML]{E0F1E7}16.4 & \cellcolor[HTML]{F3F9F8}9.6  & \cellcolor[HTML]{C8E7D2}25.4 & \cellcolor[HTML]{CAE8D4}24.4 & \cellcolor[HTML]{ABDBB9}35.8 \\
Wilcoxon 6                                                           & \cellcolor[HTML]{FCD8DB}16.3 & \cellcolor[HTML]{FCE7E9}12.6 & \cellcolor[HTML]{FCDDE0}15.1 & \cellcolor[HTML]{C2E5CD}27.6 & \cellcolor[HTML]{E9F4EE}13.5 & \cellcolor[HTML]{FBFCFE}6.8  & \cellcolor[HTML]{D2EBDB}21.8 & \cellcolor[HTML]{D3ECDC}21.2 & \cellcolor[HTML]{B5E0C2}32.1 \\
Wilcoxon 7                                                           & \cellcolor[HTML]{FAA6A8}29.4 & \cellcolor[HTML]{FBB9BB}24.4 & \cellcolor[HTML]{FBBABC}24.2 & \cellcolor[HTML]{92D1A3}44.8 & \cellcolor[HTML]{C5E6D0}26.2 & \cellcolor[HTML]{EAF5F0}13.0 & \cellcolor[HTML]{ABDBB9}35.9 & \cellcolor[HTML]{AFDDBC}34.5 & \cellcolor[HTML]{84CC98}49.7 \\
Wilcoxon 8                                                           & \cellcolor[HTML]{F98A8D}36.4 & \cellcolor[HTML]{FA9DA0}31.5 & \cellcolor[HTML]{FA9799}33.2 & \cellcolor[HTML]{80CA94}51.2 & \cellcolor[HTML]{B6E0C2}31.9 & \cellcolor[HTML]{D1EBDA}22.1 & \cellcolor[HTML]{9BD5AB}41.7 & \cellcolor[HTML]{9CD6AD}41.0 & \cellcolor[HTML]{76C68B}55.0 \\
Wilcoxon 9                                                           & \cellcolor[HTML]{F8696B}44.9 & \cellcolor[HTML]{F97C7E}40.2 & \cellcolor[HTML]{F96F71}43.6 & \cellcolor[HTML]{6AC181}59.3 & \cellcolor[HTML]{A0D7B0}39.6 & \cellcolor[HTML]{B4DFC1}32.4 & \cellcolor[HTML]{87CD9A}48.7 & \cellcolor[HTML]{88CD9B}48.2 & \cellcolor[HTML]{63BE7B}61.6 \\
Wilcoxon 10                                                          & \cellcolor[HTML]{F96C6E}44.2 & \cellcolor[HTML]{F97F82}39.3 & \cellcolor[HTML]{F97476}42.2 & \cellcolor[HTML]{6CC283}58.5 & \cellcolor[HTML]{A4D9B3}38.4 & \cellcolor[HTML]{B9E1C5}30.8 & \cellcolor[HTML]{8ACE9C}47.8 & \cellcolor[HTML]{8ACE9D}47.6 & \cellcolor[HTML]{66C07E}60.5 \\
Dietterich 5                                                         & \cellcolor[HTML]{D7EDDF}5.2  & \cellcolor[HTML]{DBEEE2}5.4  & \cellcolor[HTML]{FCF9FC}7.9  & \cellcolor[HTML]{FBF7FA}6.1  & \cellcolor[HTML]{F9A2A5}2.5  & \cellcolor[HTML]{FAB5B7}3.3  & \cellcolor[HTML]{FBE7EA}5.4  & \cellcolor[HTML]{FBFCFE}6.8  & \cellcolor[HTML]{FAC9CC}4.1  \\
Dietterich 6                                                         & \cellcolor[HTML]{E8F4EE}6.0  & \cellcolor[HTML]{F0F7F5}6.4  & \cellcolor[HTML]{FCF4F7}9.0  & \cellcolor[HTML]{FAFCFD}7.1  & \cellcolor[HTML]{FAB4B6}3.2  & \cellcolor[HTML]{FAC0C3}3.7  & \cellcolor[HTML]{FBF8FB}6.1  & \cellcolor[HTML]{F8FBFC}7.9  & \cellcolor[HTML]{FBDFE1}5.0  \\
Dietterich 7                                                         & \cellcolor[HTML]{F2F8F6}6.4  & \cellcolor[HTML]{FBFBFE}6.9  & \cellcolor[HTML]{FCF2F5}9.5  & \cellcolor[HTML]{F9FBFC}7.6  & \cellcolor[HTML]{FABEC1}3.6  & \cellcolor[HTML]{FAC9CC}4.1  & \cellcolor[HTML]{FBFCFF}6.7  & \cellcolor[HTML]{F6FAFA}8.7  & \cellcolor[HTML]{FBEDF0}5.6  \\
Dietterich 8                                                         & \cellcolor[HTML]{FCFCFF}6.9  & \cellcolor[HTML]{FCFBFE}7.3  & \cellcolor[HTML]{FCF0F3}10.0 & \cellcolor[HTML]{F7FAFB}8.1  & \cellcolor[HTML]{FAC8CB}4.1  & \cellcolor[HTML]{FAD2D4}4.5  & \cellcolor[HTML]{FAFCFE}7.1  & \cellcolor[HTML]{F4F9F8}9.3  & \cellcolor[HTML]{FBFAFD}6.2  \\
Dietterich 9                                                         & \cellcolor[HTML]{FCFBFE}7.2  & \cellcolor[HTML]{FCF9FC}7.7  & \cellcolor[HTML]{FCEFF2}10.4 & \cellcolor[HTML]{F7FAFA}8.4  & \cellcolor[HTML]{FACDD0}4.3  & \cellcolor[HTML]{FBD8DB}4.7  & \cellcolor[HTML]{F9FBFD}7.4  & \cellcolor[HTML]{F3F9F7}9.8  & \cellcolor[HTML]{FCFCFF}6.5  \\
Dietterich 10                                                        & \cellcolor[HTML]{FCFAFD}7.5  & \cellcolor[HTML]{FCF8FB}7.9  & \cellcolor[HTML]{FCEEF1}10.7 & \cellcolor[HTML]{F6FAFA}8.7  & \cellcolor[HTML]{FAD7D9}4.7  & \cellcolor[HTML]{FBDEE1}5.0  & \cellcolor[HTML]{F8FBFC}7.7  & \cellcolor[HTML]{F2F8F6}10.2 & \cellcolor[HTML]{FBFCFE}7.0  \\
Alpaydin 5                                                           & \cellcolor[HTML]{9DD5AD}2.6  & \cellcolor[HTML]{96D2A7}2.3  & \cellcolor[HTML]{85CC99}1.6  & \cellcolor[HTML]{FBDADD}4.9  & \cellcolor[HTML]{F9A5A8}2.6  & \cellcolor[HTML]{F86D6F}0.2  & \cellcolor[HTML]{F8898B}1.4  & \cellcolor[HTML]{FAB6B9}3.3  & \cellcolor[HTML]{FACDD0}4.3  \\
Alpaydin 6                                                           & \cellcolor[HTML]{A3D8B2}2.9  & \cellcolor[HTML]{98D3A9}2.4  & \cellcolor[HTML]{8ACD9C}1.8  & \cellcolor[HTML]{FBEDF0}5.6  & \cellcolor[HTML]{F9ADAF}2.9  & \cellcolor[HTML]{F86C6E}0.2  & \cellcolor[HTML]{F88B8D}1.5  & \cellcolor[HTML]{FAC4C7}3.9  & \cellcolor[HTML]{FBE5E8}5.3  \\
Alpaydin 7                                                           & \cellcolor[HTML]{A9DAB7}3.2  & \cellcolor[HTML]{9CD5AC}2.6  & \cellcolor[HTML]{8ECFA0}1.9  & \cellcolor[HTML]{FCFCFF}6.3  & \cellcolor[HTML]{F9B2B4}3.1  & \cellcolor[HTML]{F86D6F}0.2  & \cellcolor[HTML]{F88D8F}1.6  & \cellcolor[HTML]{FAD0D3}4.4  & \cellcolor[HTML]{FBF4F7}5.9  \\
Alpaydin 8                                                           & \cellcolor[HTML]{B3DEC0}3.6  & \cellcolor[HTML]{A4D8B3}2.9  & \cellcolor[HTML]{93D1A5}2.2  & \cellcolor[HTML]{FAFBFD}7.2  & \cellcolor[HTML]{FABDC0}3.6  & \cellcolor[HTML]{F86D70}0.2  & \cellcolor[HTML]{F99193}1.7  & \cellcolor[HTML]{FBDDE0}5.0  & \cellcolor[HTML]{FBFCFE}6.9  \\
Alpaydin 9                                                           & \cellcolor[HTML]{BDE2C9}4.1  & \cellcolor[HTML]{AEDCBC}3.4  & \cellcolor[HTML]{9CD5AC}2.6  & \cellcolor[HTML]{F7FAFB}8.2  & \cellcolor[HTML]{FAC9CC}4.1  & \cellcolor[HTML]{F87073}0.3  & \cellcolor[HTML]{F99597}1.9  & \cellcolor[HTML]{FBEBEE}5.6  & \cellcolor[HTML]{F8FBFC}7.9  \\
Alpaydin 10                                                          & \cellcolor[HTML]{C6E6D0}4.5  & \cellcolor[HTML]{B3DEC0}3.6  & \cellcolor[HTML]{9FD6AF}2.7  & \cellcolor[HTML]{F5FAF9}8.9  & \cellcolor[HTML]{FACFD2}4.4  & \cellcolor[HTML]{F86F71}0.3  & \cellcolor[HTML]{F99A9C}2.1  & \cellcolor[HTML]{FBF7FA}6.1  & \cellcolor[HTML]{F5FAF9}8.8
\end{tabular}
\end{table}
\clearpage

\subsection{Selection Criteria}

Here, we have listed some aspects according to which the data can be assessed. We have ranked the methods according to each aspect, and then aggregated the rankings using the Borda count.
\\[\baselineskip]
\noindent\textbf{Discriminative power (DISC):} We have taken the absolute difference between the rejection rate in the RT II and RT I scenarios ($|\text{RT I}-\text{RT II}|$); the underlying idea is that the number of folds largely explains the rejection rate of the methods. The amount a method rejects RT II instances more often than RT I instances, shows how well it can distinguish the two situations. We had no prior knowledge about the distribution of the rankings, hence we used the random scenarios. The larger the discriminative power the better.
\\[\baselineskip]
\noindent\textbf{Maximum distance from the best option (MAXDIFF):} In each scenario, we identified the best option, that is, the CV method with the best rejection rate (lowest for type I/highest for type II scenarios). Then, for each method, we measured the difference between the method's performance in the scenario and the performance of the best option. Finally, a method was evaluated by the maximum of the differences across all scenarios. Small values indicate good performance.
\\[\baselineskip]
\noindent\textbf{Average distance from the best option (AVGDIFF):} Same as the previous, with the exception that we calculated the average difference (instead of the maximum) from the best option. Again small values are preferred.
\\[\baselineskip]
\noindent\textbf{Balancedness (BLNC):}  We measured the absolute difference of type I and type II errors using the RT I and RT II columns ($|\text{RT I}-(1-\text{RT II})|$). The idea behind this is that we want to balance the errors – the smaller the difference, the better. Balancing the two types of error rates is a common method in biometrics, called crossover (or equal) error rate, see \eg\ ref.~\citep[][Chapter 5]{conradChapterDomainIdentity2017}. The smaller, the better.
\\[\baselineskip]
\noindent\textbf{Sum of Ranking Differences (SRD):}  We took the transpose of Tables~\ref{tab:results_n32}–\ref{tab:results_n7}, hence the CV methods correspond to the columns (solutions), and the scenarios to the rows (objects). Reference is the row minimum in type I scenarios (the lowest rejection rate) and the row maximum in type II scenarios (the highest rejection rate). SRD calculates how far each CV method falls from the reference. Small SRD values indicate that the method is close to the desired reference.
\\[\baselineskip]
\noindent\textbf{Pair-wise correlation methods (CEPWAVG/WTPWAVG):} Similarly to SRD, these methods work with the transposed data matrix \citep{Rajko2003}. Two solutions (X1 and X2) were selected and checked to determine whether they were related to the reference (here the average), whether both of their difference was positive (A); one of them was positive and the other was negative (B), or \emph{vice versa} (C). The frequencies were counted for all possible pairs of scenarios. Then, two statistical tests--–the conditional exact Fisher’s test (CE), and Williams’ $\td$-test (Wt)-–-decide whether the frequencies of events B and C are significantly different or not; \emph{i.e.,}\ one solution (say X1) is overriding X2, conversely, X1 loses against X2, or no decision can be made (tie). After that, the solutions were compared pairwise with the reference, considering all possible combinations (W05, W06,\dots, and A10). The solutions were further ranked according to the number of wins minus the number of losses, but the ranking was adjusted in the present case: probability-weighted ranking (pW) was used, \emph{i.e.}\ based on p(wins)-p(losses) scores. For both methods larger values indicate good performance.
\\[\baselineskip]
There are other possible aspects in terms of which the CV methods can be compared. However, since the analysis already contains seven different decision criteria, a new aspect has a small chance to turn over the aggregated ranking. As for the aggregation method, there are certainly more sophisticated ways to evaluate the data. The advantage of the Borda count is its conceptual simplicity. As Tables~\ref{tab:Borda_n32}–\ref{tab:Borda_n7} show, the results are fortunately rather straightforward.

\begin{table}[]
\scriptsize
\caption{Ranking of CV-methods in the $n=32$ case}\label{tab:Borda_n32}
\begin{tabular}{lllllllll}
              & DISCPOW & MAXDIFF & AVGDIFF & BLNC & SRD  & CEPWAVG & WTPWAVG & Borda                        \\
Wilcoxon 5    & 4       & 1       & 6       & 1    & 3.5  & 5.5     & 2       & \cellcolor[HTML]{63BE7B}103  \\
Wilcoxon 6    & 2       & 2       & 5       & 2    & 3.5  & 5.5     & 3       & \cellcolor[HTML]{63BE7B}103  \\
Wilcoxon 7    & 1       & 3       & 4       & 3    & 3.5  & 2.5     & 6       & \cellcolor[HTML]{63BE7B}103  \\
Wilcoxon 8    & 3       & 4       & 3       & 4    & 3.5  & 2.5     & 11      & \cellcolor[HTML]{7DC992}95   \\
Wilcoxon 9    & 5       & 5       & 2       & 5    & 3.5  & 2.5     & 16      & \cellcolor[HTML]{97D3A8}87   \\
Wilcoxon 10   & 6       & 6       & 1       & 6    & 3.5  & 2.5     & 17      & \cellcolor[HTML]{A1D7B1}84   \\
Dietterich 5  & 18      & 18      & 18      & 18   & 15.5 & 16      & 1       & \cellcolor[HTML]{F87173}21.5 \\
Dietterich 6  & 17      & 17      & 17      & 17   & 15.5 & 16      & 4       & \cellcolor[HTML]{F87577}22.5 \\
Dietterich 7  & 16      & 16      & 16      & 16   & 15.5 & 10.5    & 7       & \cellcolor[HTML]{F98F91}29.0 \\
Dietterich 8  & 15      & 15      & 15      & 15   & 15.5 & 16      & 15      & \cellcolor[HTML]{F8696B}19.5 \\
Dietterich 9  & 14      & 14      & 14      & 14   & 10.5 & 9       & 13      & \cellcolor[HTML]{F9B1B4}37.5 \\
Dietterich 10 & 13      & 12      & 13      & 13   & 10.5 & 10.5    & 18      & \cellcolor[HTML]{F9ABAE}36.0 \\
Alpaydin 5    & 12      & 13      & 12      & 12   & 7.5  & 7.5     & 5       & \cellcolor[HTML]{F8FBFC}57   \\
Alpaydin 6    & 11      & 11      & 11      & 11   & 7.5  & 7.5     & 8       & \cellcolor[HTML]{F2F8F6}59   \\
Alpaydin 7    & 10      & 10      & 10      & 10   & 10.5 & 12.5    & 14      & \cellcolor[HTML]{FBE0E3}49   \\
Alpaydin 8    & 9       & 9       & 9       & 9    & 10.5 & 12.5    & 12      & \cellcolor[HTML]{FBF8FB}55   \\
Alpaydin 9    & 8       & 8       & 8       & 8    & 15.5 & 16      & 9       & \cellcolor[HTML]{FBF2F5}53.5 \\
Alpaydin 10   & 7       & 7       & 7       & 7    & 15.5 & 16      & 10      & \cellcolor[HTML]{FAFCFD}56.5
\end{tabular}
\caption{Ranking of CV-methods in the $n=13$ case}\label{tab:Borda_n13}
\begin{tabular}{lllllllll}
              & DISCPOW & MAXDIFF & AVGDIFF & BLNC & SRD  & CEPWAVG & WTPWAVG & Borda                         \\
Wilcoxon 5    & 4       & 2       & 5       & 2    & 3    & 2.5     & 4       & \cellcolor[HTML]{75C58A}103.5 \\
Wilcoxon 6    & 3       & 1       & 6       & 4    & 6    & 6       & 5       & \cellcolor[HTML]{8DCF9F}95.0  \\
Wilcoxon 7    & 2       & 4       & 3       & 3    & 3    & 2.5     & 1       & \cellcolor[HTML]{69C180}107.5 \\
Wilcoxon 8    & 1       & 3       & 4       & 1    & 3    & 2.5     & 2       & \cellcolor[HTML]{63BE7B}109.5 \\
Wilcoxon 9    & 6       & 5       & 2       & 5    & 3    & 2.5     & 3       & \cellcolor[HTML]{80CA94}99.5  \\
Wilcoxon 10   & 5       & 6       & 1       & 6    & 3    & 5       & 6       & \cellcolor[HTML]{90D0A2}94.0  \\
Dietterich 5  & 18      & 18      & 18      & 17   & 15.5 & 15.5    & 9       & \cellcolor[HTML]{F8696B}15    \\
Dietterich 6  & 17      & 16      & 17      & 16   & 15.5 & 15.5    & 10      & \cellcolor[HTML]{F87779}19    \\
Dietterich 7  & 16      & 15      & 16      & 14   & 15.5 & 15.5    & 12      & \cellcolor[HTML]{F88284}22    \\
Dietterich 8  & 15      & 14      & 15      & 13   & 15.5 & 15.5    & 14      & \cellcolor[HTML]{F8898B}24    \\
Dietterich 9  & 14      & 13      & 13      & 11   & 15.5 & 15.5    & 16      & \cellcolor[HTML]{F99799}28    \\
Dietterich 10 & 13      & 12      & 12      & 10   & 15.5 & 15.5    & 18      & \cellcolor[HTML]{F99EA1}30    \\
Alpaydin 5    & 12      & 17      & 14      & 18   & 9.5  & 9.5     & 7       & \cellcolor[HTML]{FABFC1}39    \\
Alpaydin 6    & 11      & 11      & 11      & 15   & 9.5  & 9.5     & 8       & \cellcolor[HTML]{FBEAEC}51    \\
Alpaydin 7    & 10      & 10      & 10      & 12   & 9.5  & 9.5     & 11      & \cellcolor[HTML]{FBF4F7}54    \\
Alpaydin 8    & 9       & 9       & 9       & 9    & 9.5  & 9.5     & 13      & \cellcolor[HTML]{F7FAFB}58    \\
Alpaydin 9    & 8       & 8       & 8       & 8    & 9.5  & 9.5     & 15      & \cellcolor[HTML]{F1F8F6}60    \\
Alpaydin 10   & 7       & 7       & 7       & 7    & 9.5  & 9.5     & 17      & \cellcolor[HTML]{EBF6F1}62
\end{tabular}
\caption{Ranking of CV-methods in the $n=7$ case}\label{tab:Borda_n7}
\begin{tabular}{lllllllll}
              & DISCPOW & MAXDIFF & AVGDIFF & BLNC & SRD  & CEPWAVG & WTPWAVG & Borda                         \\
Wilcoxon 5    & 5       & 2       & 5       & 5    & 3    & 4       & 5       & \cellcolor[HTML]{78C78D}97.0  \\
Wilcoxon 6    & 6       & 3       & 9       & 6    & 3    & 4       & 4       & \cellcolor[HTML]{87CD9A}91.0  \\
Wilcoxon 7    & 1       & 1       & 4       & 4    & 1    & 6       & 6       & \cellcolor[HTML]{69C180}103.0 \\
Wilcoxon 8    & 2       & 4       & 3       & 3    & 3    & 4       & 2       & \cellcolor[HTML]{63BE7B}105.0 \\
Wilcoxon 9    & 4       & 11      & 1       & 2    & 5.5  & 1.5     & 3       & \cellcolor[HTML]{75C68B}98.0  \\
Wilcoxon 10   & 3       & 10      & 2       & 1    & 5.5  & 1.5     & 1       & \cellcolor[HTML]{6BC282}102.0 \\
Dietterich 5  & 15      & 12      & 12      & 12   & 15.5 & 15.5    & 18      & \cellcolor[HTML]{F87072}26.0  \\
Dietterich 6  & 11      & 9       & 11      & 11   & 15.5 & 15.5    & 16      & \cellcolor[HTML]{FAC1C3}37.0  \\
Dietterich 7  & 12      & 8       & 10      & 10   & 15.5 & 15.5    & 17      & \cellcolor[HTML]{FAC8CB}38.0  \\
Dietterich 8  & 13      & 7       & 8       & 9    & 15.5 & 15.5    & 13      & \cellcolor[HTML]{FCFCFF}45.0  \\
Dietterich 9  & 14      & 6       & 7       & 8    & 15.5 & 15.5    & 15      & \cellcolor[HTML]{FCFCFF}45.0  \\
Dietterich 10 & 16      & 5       & 6       & 7    & 15.5 & 15.5    & 14      & \cellcolor[HTML]{F7FAFB}47.0  \\
Alpaydin 5    & 18      & 18      & 18      & 18   & 9.5  & 8.5     & 11      & \cellcolor[HTML]{F8696B}25.0  \\
Alpaydin 6    & 17      & 17      & 17      & 17   & 9.5  & 11.5    & 10      & \cellcolor[HTML]{F87779}27.0  \\
Alpaydin 7    & 10      & 16      & 16      & 16   & 9.5  & 8.5     & 8       & \cellcolor[HTML]{FBE5E8}42.0  \\
Alpaydin 8    & 9       & 15      & 15      & 15   & 9.5  & 8.5     & 9       & \cellcolor[HTML]{FCFCFF}45.0  \\
Alpaydin 9    & 8       & 14      & 14      & 14   & 9.5  & 11.5    & 12      & \cellcolor[HTML]{FBEDF0}43.0  \\
Alpaydin 10   & 7       & 13      & 13      & 13   & 9.5  & 8.5     & 7       & \cellcolor[HTML]{E3F2E9}55.0
\end{tabular}
\end{table}

\subsection{Comparing Selection Criteria}\label{sec:comp_sel_crit}

\begin{figure}[!ht]
    \centering
    \includegraphics{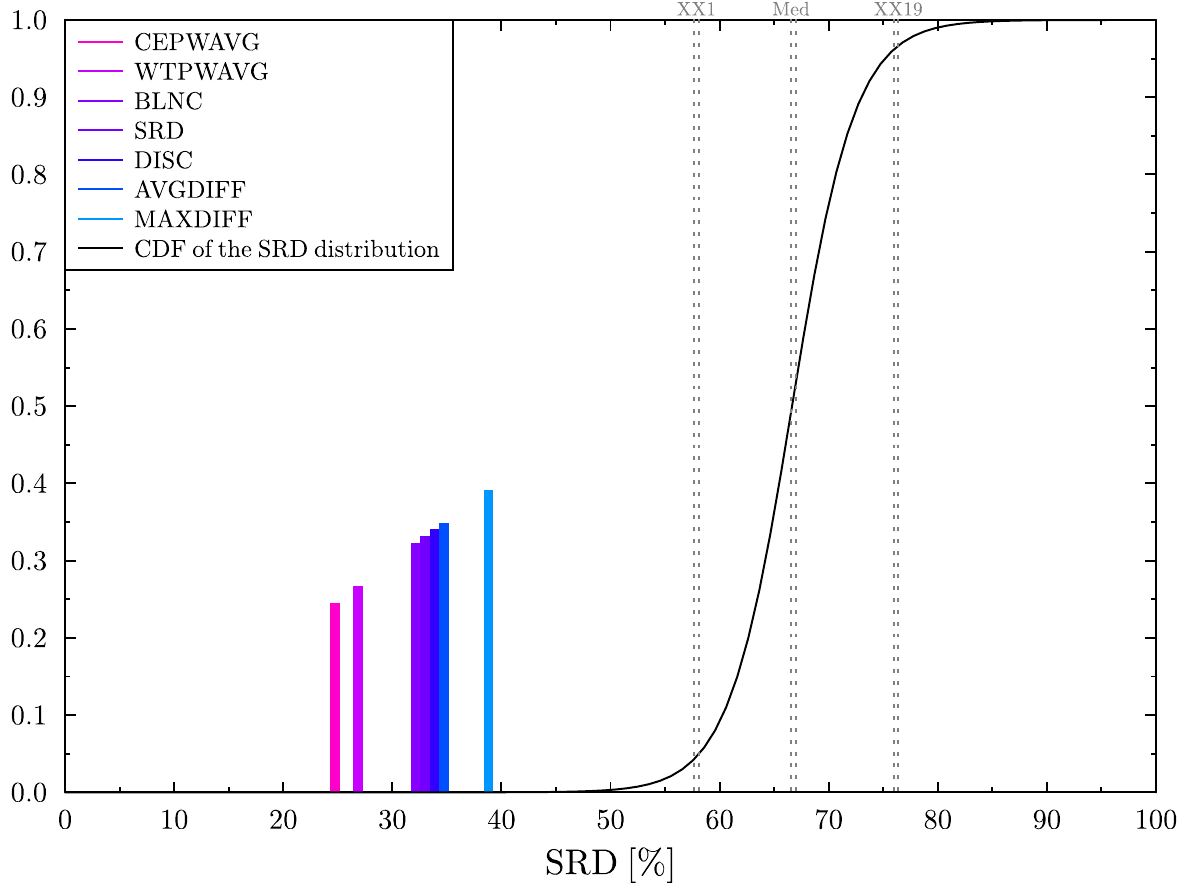}
    \caption{\textbf{Comparison of ranks with random numbers}  For sake of visualization, the colored bars' height is equal to their normalized SRD values and they follow the same order as in the legend. The black curve is a continuous approximation of the cumulative distribution function of the random SRD values. All (normalized) SRD values fall outside the 5$\%$ threshold (XX1: 5$\%$ threshold, Med: Median, XX19: 95$\%$ threshold).}\label{fig:SRD_of_sel_crit}
\end{figure}

Before we proceed with discussing the results, it is worthwhile to look at how the chosen selection criteria relate to each other. To uncover this, we performed a (meta-)SRD analysis using the data of all three input sizes. The average of the normalized performance values was taken as a reference. The idea is that using seven measures we can invoke the \say{wisdom of the crowd}: Not only the random errors but the systematic ones (biases) of different methods follow a normal distribution. If the average is used as the reference, it is assumed that the errors cancel each other out, supported by the maximum likelihood principle and empirical evidence. 
Figure~\ref{fig:SRD_of_sel_crit} displays the results.

The pair-wise correlation methods turned out to be the closest to the average evaluation of the selection criteria, followed by Balancedness and SRD itself.  The Max.\ difference from the best option was the least similar to the average, meaning that it ranked the methods according to a different dimension than the other criteria.

Notice that even the best criterion is far from the reference, indicating that  none of the methods capture every aspect of the rankings. On the other hand, all criteria fell outside the 5$\%$ threshold, that is, all of them are different from random ranking. This indicates that there is a consensus among them about the big picture.

\subsection{Model validation}\label{sec:Model_validation}

The axiomatic approach is popular in Economics. Due to its success, it has already spread to related fields, such as computational social choice.  However, to our knowledge, it has never been applied in a setting as ours. In this section, we show that the proposed scenarios are meaningful and grasp the behavior of real data.

For demonstration, we examined two unrelated datasets: laboratory performances in a quality control program to test type I situations and Elo-scores from a chess championship to test type II situations.

\subsubsection{OIL dataset}

To verify the confidence of their analytical methods, laboratories participate in a comparison program, where they have to determine some characteristics of a homogeneous sample under documented conditions. In our example, Polycyclic Aromatic Hydrocarbon contents in 16 edible oil samples (OIL) were reported by each participating laboratory \citep{Skrbic2013}. Since the laboratories work with the same substances, the expectation is that within a small statistical error they report the same measurements. Reference values were provided by the European Union
Reference Laboratory for PAHs in food (EU-RL-PAH).

From the 15 laboratories, we selected four that have shown very little discrepancy from the reference. Indeed, the various statistical tests applied in \citep{Skrbic2013} failed to show any significant difference in the measured values of these four laboratories. Thus, both expectation and empirical evidence point towards that the reported measures come from the same distribution. Clearly, this is a type I situation, where CV methods should accept the null hypothesis.

CV has a stochastic element, in each run different rows are selected. Thus, re-running a CV method on the same data may yield different result. To uncover the characteristic rejection rate on the OIL dataset, we ran each CV method 100~000 times for each pair of laboratories. Table~\ref{tab:OIL} compiles the resulting rejection rates.

\begin{table}[]
\centering
\caption{Cross-validation results for the OIL dataset. The measurements of four laboratories (L1,L4,L10,L11) were compared. Each cell shows average rejection rate ($\%$) in a sample of 100~000 runs. Last column shows the average of the pairwise comparisons.}\label{tab:OIL}
\begin{tabular}{lccccccc}
              & L1-L4 & L1-L10 & L1-L11 & L4-L10 & L4-L11 & L10-L11 & Avg                          \\
Wilcoxon 5    & 3.6   & 2.6    & 3.6    & 8.0    & 0.5    & 8.8     & \cellcolor[HTML]{FBD0D2}4.5  \\
Wilcoxon 6    & 2.7   & 1.6    & 2.7    & 9.5    & 0.1    & 9.8     & \cellcolor[HTML]{FBD1D4}4.4  \\
Wilcoxon 7    & 1.6   & 2.0    & 1.6    & 7.0    & 0.0    & 9.3     & \cellcolor[HTML]{FCDBDE}3.6  \\
Wilcoxon 8    & 2.4   & 4.1    & 2.4    & 12.5   & 0.0    & 18.4    & \cellcolor[HTML]{FBB6B9}6.6  \\
Wilcoxon 9    & 8.5   & 5.0    & 8.5    & 24.3   & 0.0    & 25.8    & \cellcolor[HTML]{F97476}12.0 \\
Wilcoxon 10   & 6.1   & 7.2    & 6.2    & 26.4   & 0.0    & 31.5    & \cellcolor[HTML]{F8696B}12.9 \\
Dietterich 5  & 1.1   & 1.1    & 0.2    & 0.2    & 1.1    & 0.3     & \cellcolor[HTML]{D1EADA}0.7  \\
Dietterich 6  & 1.7   & 1.7    & 0.3    & 0.3    & 1.8    & 0.3     & \cellcolor[HTML]{FCFBFE}1.0  \\
Dietterich 7  & 1.1   & 1.1    & 0.2    & 0.2    & 1.2    & 0.4     & \cellcolor[HTML]{DEEFE5}0.7  \\
Dietterich 8  & 1.5   & 1.6    & 0.3    & 0.3    & 1.7    & 0.4     & \cellcolor[HTML]{FCFCFF}0.9  \\
Dietterich 9  & 1.3   & 1.3    & 0.2    & 0.2    & 1.4    & 0.4     & \cellcolor[HTML]{EFF6F4}0.8  \\
Dietterich 10 & 1.5   & 1.5    & 0.2    & 0.2    & 1.7    & 0.4     & \cellcolor[HTML]{FCFCFF}0.9  \\
Alpaydin 5    & 0.8   & 0.1    & 0.8    & 0.0    & 0.7    & 0.0     & \cellcolor[HTML]{9ED6AE}0.4  \\
Alpaydin 6    & 0.5   & 0.0    & 0.5    & 0.0    & 0.5    & 0.0     & \cellcolor[HTML]{85CB98}0.3  \\
Alpaydin 7    & 0.3   & 0.0    & 0.3    & 0.0    & 0.3    & 0.0     & \cellcolor[HTML]{70C386}0.2  \\
Alpaydin 8    & 0.2   & 0.0    & 0.2    & 0.0    & 0.2    & 0.0     & \cellcolor[HTML]{63BE7B}0.1  \\
Alpaydin 9    & 0.3   & 0.0    & 0.3    & 0.0    & 0.3    & 0.0     & \cellcolor[HTML]{6FC285}0.1  \\
Alpaydin 10   & 0.2   & 0.0    & 0.2    & 0.0    & 0.1    & 0.0     & \cellcolor[HTML]{63BE7B}0.1
\end{tabular}
\end{table}

Sixteen substances are compared in the OIL dataset, with the previous notation, $n=16$. Given the small differences between the measured values, it is if we applied 8 or 10 random inversions on the reference.  The closest scenario in our axiomatic analysis is $13|13$ under the $n=13$ case. The result (Avg column in Table~\ref{tab:OIL}) indeed resembles what we see in Table~\ref{tab:results_n13}. All methods managed to improve their efficiency, most notably Wilcoxon 5 to 8 is very close to the desired 5$\%$ threshold. In comparison, the rejection rate of the Wilcoxon variants ranged between 25$\%$ and 40$\%$ during the simulations under the artificial scenario.

\subsubsection{Chess dataset}

There is a long tradition in chess to measure the performance of the players. The Elo score system quantifies the playing strengths. The difference between Elo ratings can be translated into winning probabilities: how likely is that the stronger player (with a higher rating) beats the weaker one (with a lower rating).

We chose the Grand Swiss tournament of 2019, and looked at the preliminary Elo ratings, tournament performance (also measured in Elo points), and post-tournament Elo ratings of the top 32 players\footnote{Data was gathered from ChessResults.com: \url{http://chess-results.com/tnr478041.aspx?lan=1&art=1&flag=30}}.

Tournament performance is calculated based on the opponents' Elo rating and match scores and it may depend on various factors, including quality of preparation and match pairings, but also on unexpected events like the mood of the players or a sudden illness. Based on the wins and losses, the post-tournament Elo rating differs from the preliminary rating by a couple of Elo points. Tournament performance, on the other hand, fluctuates wildly.

\cite{sziklai2021efficacy} compare the performances of 900 players in six different tournaments and derive an empirical distribution of the Elo rating difference. The distribution resembles a Gaussian curve, but with a strong tail, meaning that extremely good or bad performances are not that uncommon.

The same is true for our data. The ranking of the 32 players based on their performances visible differs from either the preliminary or the post-tournament rankings. Preliminary rankings are based on the match history of the players. The preliminary Elo scores might not be up-to-date since it could have passed months since the last rated game was played by the contestants. Although as we mentioned, there are a plethora of reasons why a chess match is decided one way or the other, most wins are based on merit. Arguably, the post-tournament ratings are the closest to the true power ranking of the players, hence we choose it as the reference. From this point of view,  preliminary ratings and tournament performances are two perturbations of different amplitude. This is clearly a type II situation, CV methods should be able to distinguish between the two data columns.

As in the previous case, we ran 100~000 simulations to uncover the characteristic rejection rates of the methods. Table~\ref{tab:CHESS} compiles the result. Since the number of players is the same as the number of rows in our first simulation setup, the results are directly comparable to the numbers in Table~\ref{tab:results_n32}. However, the difference between the rankings is much larger. The most powerful transformation we applied in the $n=32$ case is 88 inversions which created on average a 0.11 normalized SRD score (\emph{cf.}\ Fig.~\ref{fig:trans_SRD}). Here, the normalized SRD distance between the preliminary ranking and the reference is 0.15, while the tournament performance differs with a hefty 0.29 SRD score. This means that the difference between the rankings is more pronounced. Still, we would expect something similar to the 64|16 scenario and indeed that is what we see in Table~\ref{tab:CHESS}. Again, the only difference is that the Wilcoxon variants perform better (with 100$\%$ efficacy!) than in the simulation.

\begin{table}[]
\centering
\caption{Cross-validation results for the CHESS dataset. Preliminary Elo ratings are cross tested with tournament performances. Each cell shows average rejection rate ($\%$) in a sample of 100~000 runs. }\label{tab:CHESS}
\begin{tabular}{lclc}
              & \multicolumn{1}{c}{prelim.\ \emph{v.} tour.\ perf.} & & \multicolumn{1}{c}{prelim.\ \emph{v.} tour.\ perf.}    \\
Wilcoxon 5   & \cellcolor[HTML]{64BF7C}99.99 & Dietterich 8  & \cellcolor[HTML]{F88486}18.31 \\
Wilcoxon 6   & \cellcolor[HTML]{64BF7C}100.00                         & Dietterich 9  & \cellcolor[HTML]{F8898B}18.79 \\
Wilcoxon 7   & \cellcolor[HTML]{64BF7C}100.00                         & Dietterich 10 & \cellcolor[HTML]{F88B8D}19.01 \\
Wilcoxon 8   & \cellcolor[HTML]{64BF7C}100.00                         & Alpaydin 5    & \cellcolor[HTML]{F88688}18.54 \\
Wilcoxon 9   & \cellcolor[HTML]{64BF7C}100.00                         & Alpaydin 6    & \cellcolor[HTML]{F9B2B4}22.64 \\
Wilcoxon 10  & \cellcolor[HTML]{64BF7C}100.00                         & Alpaydin 7    & \cellcolor[HTML]{FBE2E5}27.24 \\
Dietterich 5 & \cellcolor[HTML]{F8696B}15.74 & Alpaydin 8    & \cellcolor[HTML]{F7FAFB}31.99 \\
Dietterich 6 & \cellcolor[HTML]{F87577}16.91 & Alpaydin 9    & \cellcolor[HTML]{ECF6F1}37.15 \\
Dietterich 7 & \cellcolor[HTML]{F87D7F}17.68 & Alpaydin 10   & \cellcolor[HTML]{E1F1E8}42.33
\end{tabular}
\end{table}

\hfill\\
To conclude, the test runs on the real data show very similar results to the simulations of the appropriate scenarios. The slight differences stem from the variation in the number of rows or the magnitude of the perturbations. By and large, the scenarios paint an accurate picture of what is expected under various circumstances.

\section{Discussion}\label{sec:Discussion}

With some rare exceptions, all selection criteria rank the six Wilcoxon variants in the first six places. The case studies show that the simulation scenarios adequately model what happens on real data. The only difference is that in both case studies Wilcoxon's performance exceeded expectations.
The downside of Wilcoxon is that it is too sensitive: it picks up on subtle differences in the data and rejects the null hypothesis even if it is ought to not.

Dietterich and Alpaydin are near the desired 5$\%$ threshold in type I situations. In type II situations Alpaydin outperforms Dietterich, although it is still very bad compared to Wilcoxon. Hence, if the practitioner is not afraid of categorizing different things as the same it may select Alpaydin CV. Among the Alpaydin variants the 10-fold version performs the best\footnote{For large $n$, Borda seems to favor Alpaydin 5-6 over 10 (see Table~\ref{tab:Borda_n32}). However, this happens because SRD and the pairwise correlation methods do not consider absolute differences. Alpaydin 5 and 6 are indeed slightly better in 64|1u and  88|4m situations, but Alpaydin 10 is much better in 64|16 cases.}. Although it is uncommon to apply a fold greater than 10, it may improve the test's performance in type II situations. The simulation results also suggest that Alpaydin's power becomes better as we increase the data size, that is, the number of rows in the analysis.

That being said, it is rarely the case that we only care about one type of error. Both Alpaydin and Dietterich perform poorly in type II situations. How poorly?

Figure~\ref{fig:SRD_64_32_16} sheds light on the expected rejection rates in the 64|16 and 32|16 scenarios. Although it is difficult to quantify the ideal rejection rates, in the 64|16 scenario a rejection rate of 90+$\%$ is definitely preferable to a rate of 20-50$\%$. Even in the 32|16 scenario, a rate of 60-80$\%$ is closer to the truth than a rate of 5-15$\%$.

Another problem with Dietterich and Alpaydin is that they are incapable to recognize 2n|1u or x|4m type of scenarios, although such structures in the data can be easily detected by the naked eye. Interestingly, Wilcoxon performs convincingly in these scenarios. One possible reason behind the difference in the performances can be the way these CV methods leave out rows. Dietterich and Alpaydin leave out half the rows in each fold, while Wilcoxon only excludes 10-20\%.

It is somewhat understandable that all CV methods struggle with the $(n/2)t|(n/2)b$ type of scenarios. Row selection is randomized for all three CV methods, thus they cannot distinguish between the first and second part of the data. Real datasets, however, are often composed of blocks so this scenario is very much relevant. The only remedy is if we encompass such lessons in the CV process and let the row selection follow some pattern instead of randomness. Note that Wilcoxon outperformed the other two methods in this aspect as well.

Let us summarize our findings.

\begin{itemize}
    \item Based on the aggregated indicators, Wilcoxon~8 performed the best, although Wilcoxon~7 was also very good. Tests with real data also support the selection of Wilcoxon.

    \item The Borda scores indicate that Alpaydin is to be preferred over Dietterich under all data sizes.

    \item Alpaydin prevails in type I situations, but only performs somewhat satisfactorily in type II scenarios if both the number of folds and the size of the data $(n)$ are large.
    \item None of the methods were particularly good. Therefore, future research is needed to discover a more efficacious combination of a statistical test and a suitable CV method.
\end{itemize}

\section{Summary and Conclusions}

We tested how successful the combinations of statistical tests and CV methods are in distinguishing rankings that come from various distributions. Despite the widespread usage of rankings, to our knowledge, this is the first paper that has tackled this problem. Our method of analysis is innovative, as we devised simulation scenarios to uncover the strengths and weaknesses of certain methods. This approach resembles axiomatic analysis, a common practice in theoretical economics, especially in social choice.

We investigated three tests, Wilcoxon's, Dietterich's, and Alpaydin's, each combined with folds ranging from 5 to 10, making 18 variants altogether (Tables~\ref{tab:results_n32}–\ref{tab:results_n7}). The simulation data was ranked according to various decision criteria, and the rankings were then aggregated to choose a winner. Wilcoxon test with 8 folds proved to be the best. Although Wilcoxon is a bit too sensitive in type I situations, it is the only method that performs well in type II scenarios, and is especially good in picking up structures in the data.  The Dietterich and Alpaydin tests fared poorly in type II situations, although Alpaydin was somewhat better than Dietterich. Alpaydin should only be chosen if the practitioner does not want to err in type I situations. In such cases, Alpaydin~10 is recommended. Tests with real data affirmed our findings.

Another interesting observation is that Alpaydin~10 dominates the 5-fold Alpaydin test in almost every aspect (Tables~\ref{tab:Borda_n32}–\ref{tab:Borda_n7}), even though in practice, the latter is used almost always. It would be interesting to see how Alpaydin behaves with an even higher fold number.


We compared CV methods in the framework of Sum of Ranking Differences. That means, that the differences between the rankings were measured by Manhattan distance or L1-norm. There are other meaningful measures, for instance, the number of inversions (Kendall tau). An interesting future research direction is to test whether the choice of distance has any effect on the results -- although we do not expect big surprises in this aspect.

Finally, there is a lot of room for improvement regarding to rejection rates. Even the best CV method was not particularly good (Section~\ref{sec:comp_sel_crit})–-there were scenarios where it performed poorly. Future research is needed to devise a more efficacious CV method to distinguish different rankings. Interestingly, both Dietterich and Alpaydin seem to improve if we adjust their calculation and use the absolute difference in the numerator $(\Delta_i)$. However, if we modify their formula they follow different distributions than what \cite{dietterichApproximateStatisticalTests1998} and \cite{alpaydinCombinedCvTest1999} proved.

\section*{Acknowledgments}
Bal\'azs R.\ Sziklai is the grantee of the J\'anos Bolyai Research Scholarship of the Hungarian Academy of Sciences. This research was supported by the National Research, Development and Innovation Office of Hungary
OTKA grants K~138945 (Sziklai) and K~134260 (H\'eberger).




\bibliography{bib}



\end{document}